\documentclass[aps,prl,twocolumn,showpacs,10pt,superscriptaddress,preprintnumbers,nofootinbib]{revtex4-1}
\usepackage{epsfig,amssymb,amsmath,psfrag,epstopdf,xcolor}
\pdfoutput=1
\usepackage{graphicx}
\usepackage{subcaption}
\usepackage{ragged2e}
\DeclareCaptionJustification{justified}{\justifying}
 \usepackage[normalem]{ulem}
\usepackage{paralist}
\usepackage{hyperref}
\allowdisplaybreaks

\def\beq{\begin{equation}}
\def\eeq{\end{equation}}
\def\bsp#1\esp{\begin{split}#1\end{split}}
\newcommand{\be}{\begin{equation}}
\newcommand{\ee}{\end{equation}}
\newcommand{\bea}{\begin{eqnarray}}
\newcommand{\eea}{\end{eqnarray}}

\def\Fig#1{Fig.~{\ref{#1}}}

\def\to{\rightarrow}

\def\dead{{\theta_{0}}}

\newcommand{\comment}[1]{}

\newcommand{\td}{\mathrm{d}}

\renewcommand{\v}[1]{\mathbf{#1}}

\renewcommand{\>}{\right\rangle}
\newcommand{\<}{\left\langle}

\newcommand{\As}{\alpha_{\mathrm{s}}}

\usepackage{bm}
\let\v=\bm

\begin{document}

%%%%%%%%%%%%%%%%%%%%%%%%%%%%%%%%%%%%%%%%%%
\title{Seeing Beauty in the Quark-Gluon Plasma with Energy Correlators}

\author{Carlota Andres}
\affiliation{CPHT, CNRS, \'Ecole polytechnique,  Institut Polytechnique de Paris, 91120 Palaiseau, France}
\affiliation{LIP, Av. Prof. Gama Pinto, 2, P-1649-003 Lisboa, Portugal}

\author{Fabio Dominguez}
\affiliation{Instituto Galego de F{\'{i}}sica de Altas Enerx{\'{i}}as (IGFAE),  Universidade de Santiago de Compostela, Santiago de Compostela 15782,  Spain}

\author{Jack Holguin}
\affiliation{CPHT, CNRS, \'Ecole polytechnique, Institut Polytechnique de Paris, 91120 Palaiseau, France}
\affiliation{University of Manchester, School of Physics and Astronomy, Manchester, M13 9PL, United Kingdom}

\author{Cyrille Marquet}
\affiliation{CPHT, CNRS, \'Ecole polytechnique,  Institut Polytechnique de Paris, 91120 Palaiseau, France}

\author{Ian Moult}
\affiliation{Department of Physics, Yale University, New Haven, CT 06511}

%%%%%%%%%%%%%%%%%%%%%%%%%%%%%%%%%%%%%%%%%%%%%%%%%%%%%%%%%%%%%%%%%%%%%
\begin{abstract}
Heavy quarks created in heavy-ion collisions serve as an excellent probe of the produced quark-gluon plasma (QGP). The radiation pattern of jets formed from heavy quarks as they traverse the QGP exhibits a particularly interesting structure due to the interplay of two competing effects: the suppression of small-angle radiation, also known as the ``dead-cone'' effect, and the enhancement of emitted gluons by medium-induced radiation. In this \emph{Letter}, we propose a new observable, based on the energy correlator approach to jet substructure, which will allow us to disentangle the two scales associated to these two phenomena and to determine under which conditions the dead-cone is filled by medium-induced radiation. Combined with the forthcoming high-statistics measurements of heavy-flavor jets, this work provides a novel tool to unravel the dynamics of the QGP.
\end{abstract}
%%%%%%%%%%%%%%%%%%%%%%%%%%%%%%%%%%%%%%%%%%%%%%%%%%%%%%%%%%%%%%%%%%%%%

\maketitle

\emph{Introduction.}--- 
Understanding Quantum Chromodynamics (QCD) under extreme conditions stands as a  major goal of modern nuclear physics. In this regard, the direct production of  the quark-gluon plasma (QGP) in heavy-ion collisions \cite{Gyulassy:2004zy,PHOBOS:2004zne,Muller:2006ee,Muller:2012zq} provides a unique opportunity to explore the dynamics of a strongly coupled relativistic quantum field theory. For reviews see~\cite{Busza:2018rrf,Dexheimer:2020zzs,Arslandok:2023utm}. By studying highly energetic partons generated in the underlying initial hard collisions, and their subsequent propagation through the QCD medium, we can probe the microscopic properties of the QGP. These partons undergo radiation within the QCD matter, resulting in the formation of hadronic jets that carry imprints of the internal dynamics of the QGP. Unraveling these correlations is the primary focus of the field of jet substructure, which has witnessed impressive progress in both theoretical and experimental aspects  over the past decade \cite{Larkoski:2017jix,Kogler:2018hem, Connors:2017ptx,Cunqueiro:2021wls}. 

Jets originating from a heavy quark, namely $c$ and $b$ quarks, are ideal probes of the properties of the QGP.  Due to their large masses, they propagate as long-lived particles through the entire evolution of the system, as opposed to light-quark jets which can rotate into gluon jets.  As a result, they have garnered significant attention in the literature,  as evidenced by numerous studies~\cite{Djordjevic:2003zk,Djordjevic:2004nq,Armesto:2003jh,Zhang:2003wk,Zhang:2004qm,Li:2018xuv,Kang:2016ofv,Xing:2021xwc,Huang:2013vaa,Kang:2018wrs,Li:2017wwc,Cao:2018ews}, as well as reviews and experimental measurements~\cite{Andronic:2015wma,Rapp:2018qla,Dong:2019byy,Apolinario:2022vzg,CMS:2018bwt,CMS:2013qak,CMS:2018dqf,CMS:2015sfx,CMS:2022btc}. While $c$- and $b$-jets are relatively rare compared to massless jets, the large datasets from present and future runs at the Relativistic Heavy Ion Collider (RHIC) and Large Hadron Collider (LHC) will enable precision studies of the substructure of heavy-flavor jets in heavy-ion collisions. This motivates a renewed attention in understanding and characterizing their radiation patterns. 

In recent years, there has been a revitalized interest in studying the ``dead-cone'' effect \cite{Dokshitzer:1991fd}, which refers to the suppression of vacuum radiation off a heavy quark below a certain angle known as the dead-cone angle:
\be
\dead \sim \frac{m_{Q}}{E}\,,
\ee
where $m_Q$ denotes the quark mass and $E$ its energy. This effect has been recently measured in p-p collisions \cite{ALICE:2021aqk} using an approach proposed in \cite{Cunqueiro:2018jbh} that employs iterative declustering techniques based on the angular ordering of collinear radiation. However, the situation becomes less clear in heavy-ion collisions, where medium-induced gluon radiation has been argued to fill the dead-cone region \cite{Armesto:2003jh}. Additionally, performing the same type of measurement as in vacuum may not be feasible in the heavy-ion environment, where medium-induced emissions do not follow angular ordering \cite{Mehtar-Tani:2010ebp} and grooming techniques are prone to misidentify splittings \cite{Mulligan:2020tim}.  Although efforts are underway to develop grooming algorithms specifically designed for this purpose \cite{Cunqueiro:2022svx}, the filling of the dead-cone in heavy-ion collisions provides an ideal setting for exploring another kind of observable: correlations of energy flow operators \cite{Basham:1979gh,Basham:1978zq,Basham:1978bw,Basham:1977iq,Hofman:2008ar}.

This approach consists of directly studying correlation functions of energy flow operators $\<\mathcal{E}(\vec{n}_{1}) \dots \mathcal{E}(\vec{n}_{k})\>$, where $\mathcal{E}(\vec{n}_1)$ measures the asymptotic energy flux in the direction $\vec{n}_1$ \cite{Sveshnikov:1995vi,Tkachov:1995kk,Korchemsky:1999kt,Bauer:2008dt,Hofman:2008ar,Belitsky:2013xxa,Belitsky:2013bja,Kravchuk:2018htv}. These correlation functions, commonly known as \textit{energy correlators}, provide a direct link between the experimentally accessible macroscopic energy flux and the microscopic parameters of the underlying field theory. Initially, the simple behavior exhibited by energy correlators in conformal field theories attracted considerable attention \cite{Hofman:2008ar}, leading to significant advancements in understanding their theoretical properties and the development of novel calculation techniques \cite{Belitsky:2013xxa,Belitsky:2013bja,Belitsky:2013ofa,Korchemsky:2015ssa,Belitsky:2014zha,Dixon:2018qgp,Luo:2019nig,Henn:2019gkr,Chen:2019bpb,Dixon:2019uzg,Korchemsky:2019nzm,Chicherin:2020azt,Kravchuk:2018htv,Kologlu:2019bco,Kologlu:2019mfz,Chang:2020qpj,Dixon:2019uzg,Chen:2020uvt,Chen:2020vvp,Chen:2019bpb,Chen:2020adz,Chicherin:2020azt,Chen:2021gdk,Korchemsky:2021okt,Korchemsky:2021htm,Chang:2022ryc,Chen:2022jhb,Chen:2022swd,Lee:2022ige,Yan:2022cye,Yang:2022tgm,Chen:2023wah}. Subsequently, it was realized that  energy correlators possess a key feature: their sensitivity to the presence of any intrinsic or emergent scales, which manifest as distinctive angular scales in their spectra. This remarkable property has lead to a multitude of recent applications in both high-energy and nuclear physics \cite{Komiske:2022enw,Holguin:2022epo,Liu:2022wop,Liu:2023aqb,Cao:2023rga,Devereaux:2023vjz,Andres:2022ovj,Andres:2023xwr,Craft:2022kdo,Lee:2022ige}. 

%%%%%%%%%%%%%%%%%%%%%%%%%%%%%%%%%%%%%%%%%%%%%%%%%%%%%%%%%%%%%%%%%%%%%
\begin{figure}[ht]
\includegraphics[width=0.40\textwidth]{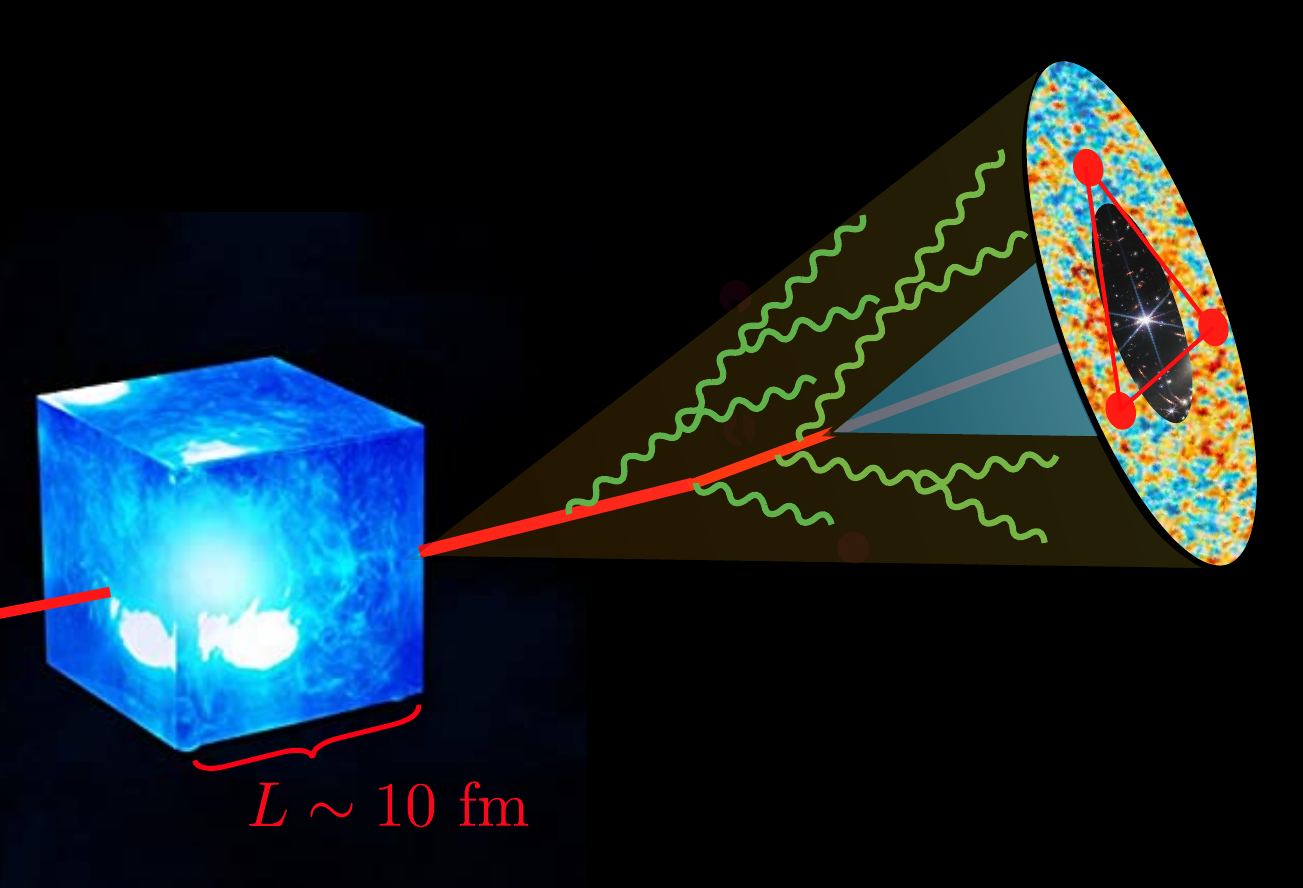}
  \caption{A heavy-flavor jet propagating through the QGP forms a complicated energy pattern due to an interplay of two characteristic angular scales: the dead-cone angle $\dead$ and the onset angle $\theta_{\rm on}$.  These scales can be extracted from the asymptotic energy flux using energy correlators.}
  \label{fig:schematic}
\end{figure}
%%%%%%%%%%%%%%%%%%%%%%%%%%%%%%%%%%%%%%%%%%%%%%%%%%%%%%%%%%%%%%%%%%%%%%

The study of jet substructure in heavy-ion collisions through energy correlators was first proposed in \cite{Andres:2022ovj,Andres:2023xwr} for jets initiated by massless quarks. This pioneering work showed that medium-induced radiation does not modify the perturbative structure of the correlator spectra at small angles yielding an enhancement above a certain scale known as the onset angle $\theta_{\rm on}$. For the case of heavy-flavor jets, we anticipate the correlators to display a richer structure  due to the interplay of two competing effects. On the one hand, vacuum radiation is suppressed below the dead-cone angle $\dead$, an effect which has been shown to result in a significant depletion of the correlator spectra for heavy-quark p-p jets below the characteristic angle $\dead$ \cite{Craft:2022kdo}. On the other hand, the correlator spectra is enhanced by medium-induced emissions above the onset angle $\theta_{\rm on}$. The interplay of these scales manifests in subtle correlations in the radiation pattern of heavy-flavor jets, which we propose to be measured using energy correlators, as illustrated in \Fig{fig:schematic}.

In this \emph{Letter}, we present the first calculation of the medium-modified two-point energy correlator $\<\mathcal{E}(\vec{n}_{1})  \mathcal{E}(\vec{n}_{2})\> $ for heavy-flavor jets, which introduces the angular scale $\cos \theta= \vec{n}_{1}\cdot \vec{n}_{2}$. Following \cite{Andres:2022ovj,Andres:2023xwr}, we adopt a simplified ``brick" model, featuring a static plasma of fixed length. Within this framework, we show the ability of the correlator spectra to disentangle the two angular scales $\dead$ and $\theta_{\rm on}$. These results provide first illustration of the extraction of two competing scales using energy correlators, highlighting in addition their remarkable potential in unraveling the intricate dynamics of the QGP.

%%%%%%%%%%%%%%%%%%%%%%%%%%%%%%%%%%%%%%%%%%%%%%%%%%%%%%%%%%%%%%%%%%%%%
\emph{Theoretical Approach.}--- We consider the two-point energy correlator (EEC) of  a heavy-quark jet with initial energy  $E$ as it traverses a QCD medium.  We assume the heavy quark to be long-lived, approximating an experimental setup where the momentum of the D or B meson is reconstructed and its decay products are removed from the jet on which the energy correlators are measured. The effects of the medium on the jet splitting with respect to the vacuum are captured in the function $F_{\rm med}$. Provided that $F_{\rm med}\rightarrow 0$ at vanishingly small angles, we can express the medium-modified EEC as \cite{Andres:2022ovj,Andres:2023xwr}
\begin{align}
    &\frac{\td \Sigma}{\td \theta} = \frac{1}{\sigma}\int \td z  \left(g^{(1)}(\theta,\As) + F_{\rm med}(z,\theta) \right) \label{eq:master} \\
    &~~~~ \times \frac{\td \hat\sigma^{\rm vac}_{qg}}{\td \theta \td z }  z (1-z) \left(1  + \mathcal{O}\left(\frac{\bar \mu_{\rm s}}{E} \right)\right) + \mathcal{O}\left(\frac{\Lambda_{\mathrm{QCD}}}{\theta \, E}\right) \,, \nonumber 
\end{align}
where $\td \hat\sigma^{\rm vac}_{qg}$ is the fixed order vacuum inclusive cross-section for a quark jet
to split into a semi-hard quark subjet and semi-hard gluon subjet, $z$ is the gluon energy fraction, and $g^{(1)}$ captures the small angle vacuum resummation. Here $\bar \mu_{\rm s}$ is the low scale of radiation over which $\td \sigma_{qg}$ is inclusive \cite{Andres:2023xwr}. We adopt the simplifying assumption of not applying any jet algorithm in our study. If a jet algorithm were to be used, the jet radius needs to be parametrically large relative to the angular size of the correlations~\cite{Lee:2022ige}. Note that by
construction the two-point correlator in p-p collisions, ${\rm d}\Sigma_{\rm vac}/{\rm d} \theta$, is achieved by setting $F_{\rm med} = 0$ in \eqref{eq:master}.

To compute the function $F_{\rm med}(z,\theta)$ one must choose a jet quenching model. Following \cite{Andres:2022ovj,Andres:2023xwr} we adopt the semi-hard approximation \cite{Dominguez:2019ges,Isaksen:2020npj} of the multiple scattering BDMPS-Z formalism \cite{Baier:1996kr,Baier:1996sk,Zakharov:1996fv,Zakharov:1997uu} in the main body of the text. This specific implementation assumes that all partons propagate along straight-line trajectories, allowing for the resummation of multiple scatterings. We restrict ourselves to the case of a static QGP of length $L$ with constant linear density of scatterings $n(t)=n_0\,\Theta(t-L)$. In this scenario, the onset angle, which parametrically indicates the minimum angle for emissions to be sensitive to medium modifications, is proportional to $\theta_L$, which is the minimal angle for which there can be emissions with a formation time smaller than the length of the medium and for a light-quark jet goes as  $\theta_L = 1/\sqrt{L E}$ \cite{Andres:2023xwr}. We model the interactions between the partons and the medium using a screened Coulomb-like interaction, commonly known as Yukawa or Gyulassy-Wang model~\cite{Gyulassy:1993hr}. The explicit expressions for $F_{\rm med}(z,\theta)$ can be found in Eq.~(3.28) and Eqs.~(3.36),~(3.37)~and~(3.38) of ref.~\cite{Andres:2023xwr}. For our numerical calculations, we set the constant screening mass $\mu$ to $\mu=1$ GeV. In our previous work \cite{Andres:2023xwr}, we compared this approach with both the harmonic oscillator and the single scattering approximations for the massless case, and found consistent results. We expect the same agreement to hold for the massive case as well. We show this agreement in Appendix~A, where we present concurrent results for the single scattering approximation.

As compared to the massless case, the leading-log mass dependence in Eq.~\eqref{eq:master} enters via
\begin{enumerate}
    \item $\td \hat\sigma^{\rm vac}_{qg} / \td \theta \td z $, which we compute at LO in the small angle limit using the massive $\mathcal{O}(\As)$ $q\rightarrow qg$ collinear splitting function \cite{Catani:2000ef}.
    \item $F_{\rm med}$, which is now computed  using massive quark propagators. In the semi-hard approximation employed in this manuscript, the mass dependence contributes to $F_{\rm med}$ only through phase factors in the vacuum propagators of eq.~(3.17) of \cite{Andres:2023xwr} where we must replace $\v{p}_1^2\to\v{p}_1^2+m_Q^2$ for the propagators of the massive quark.\footnote{In the derivation of $F_{\rm med}$, as explained in Appendix A of \cite{Isaksen:2020npj}, one should also modify the vertex factor coming from the Dirac algebra of the vacuum emission. This modification is absorbed into the mass correction to the vacuum splitting function (point 1) and so does not change $F_{\rm med}$.} As a result, the calculation of $F_{\rm med}$ can be performed using the massless derivation in \cite{Andres:2023xwr}, with the vacuum formation time replaced by its corresponding massive version, given by
    \beq
    t_{\rm f} = \frac{2}{z(1-z)E\left(\theta^2 + \frac{m_Q^2}{(1-z)^2E^2}\right)}\,. 
    \eeq
\end{enumerate}

We note that in \eqref{eq:master} we have neglected energy loss effects, since they are suppressed for large jet radii due to the inclusivity of the correlator. However, for the process considered here, energy loss can still induce a bias on the energy of the heavy-ion jet, leading to a shift in the A-A EEC toward smaller angles. This bias was not present in previous studies of energy correlators in heavy-ion collisions, since the jet could be tagged with a vector boson \cite{Andres:2022ovj,Andres:2023xwr}. It would be interesting to study this more quantitatively, see \cite{Andres:2023xwr,Barata:2023vnl} for recent progress.
%

%%%%%%%%%%%%%%%%%%%%%%%%%%%%%%%%%%%%%%%%%%%%%%%%%%%%%%%%%%%%%%%%%%%%%%
\begin{figure}[ht]
\includegraphics[width=0.45\textwidth]{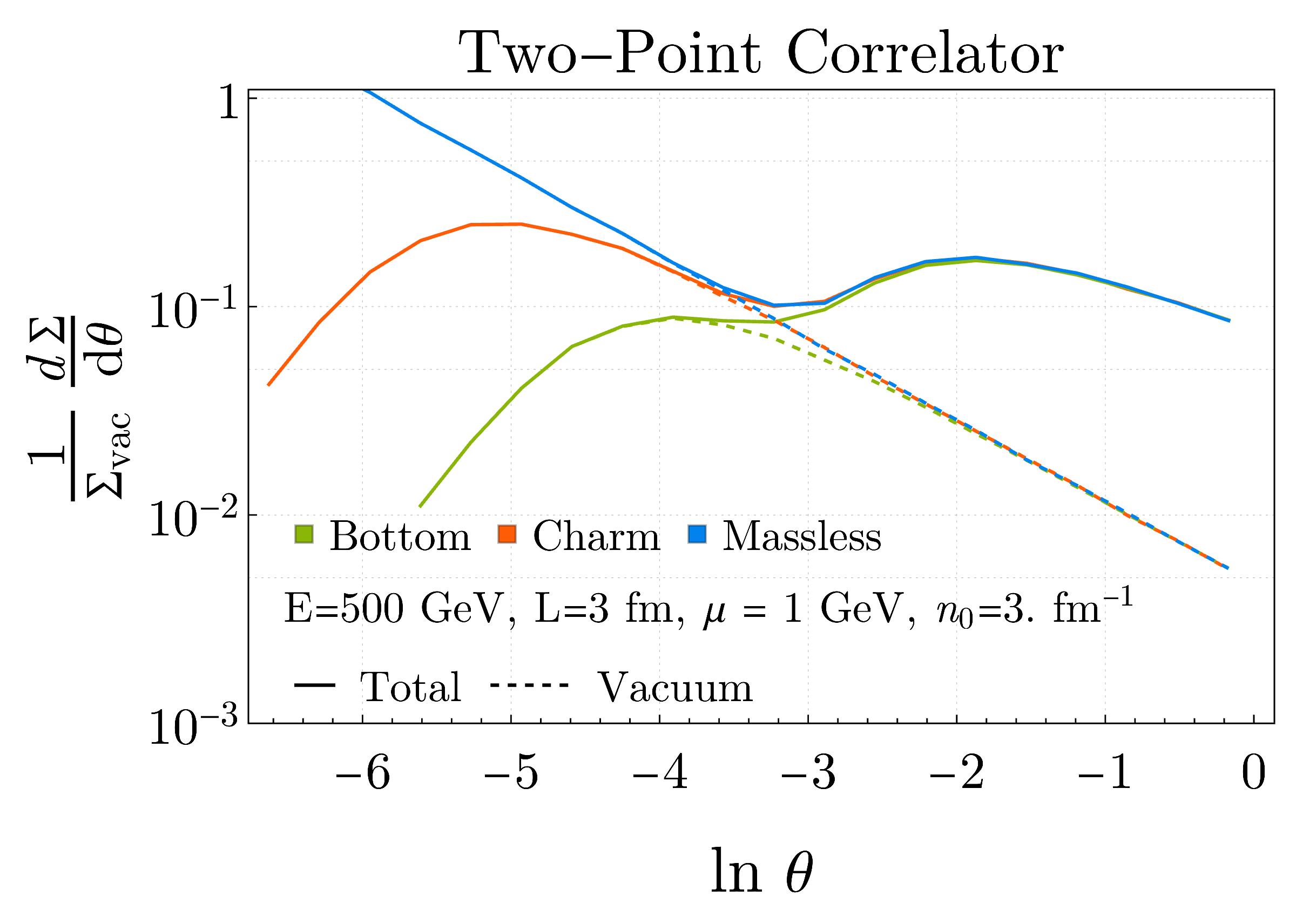}\\
\includegraphics[width=0.45\textwidth]{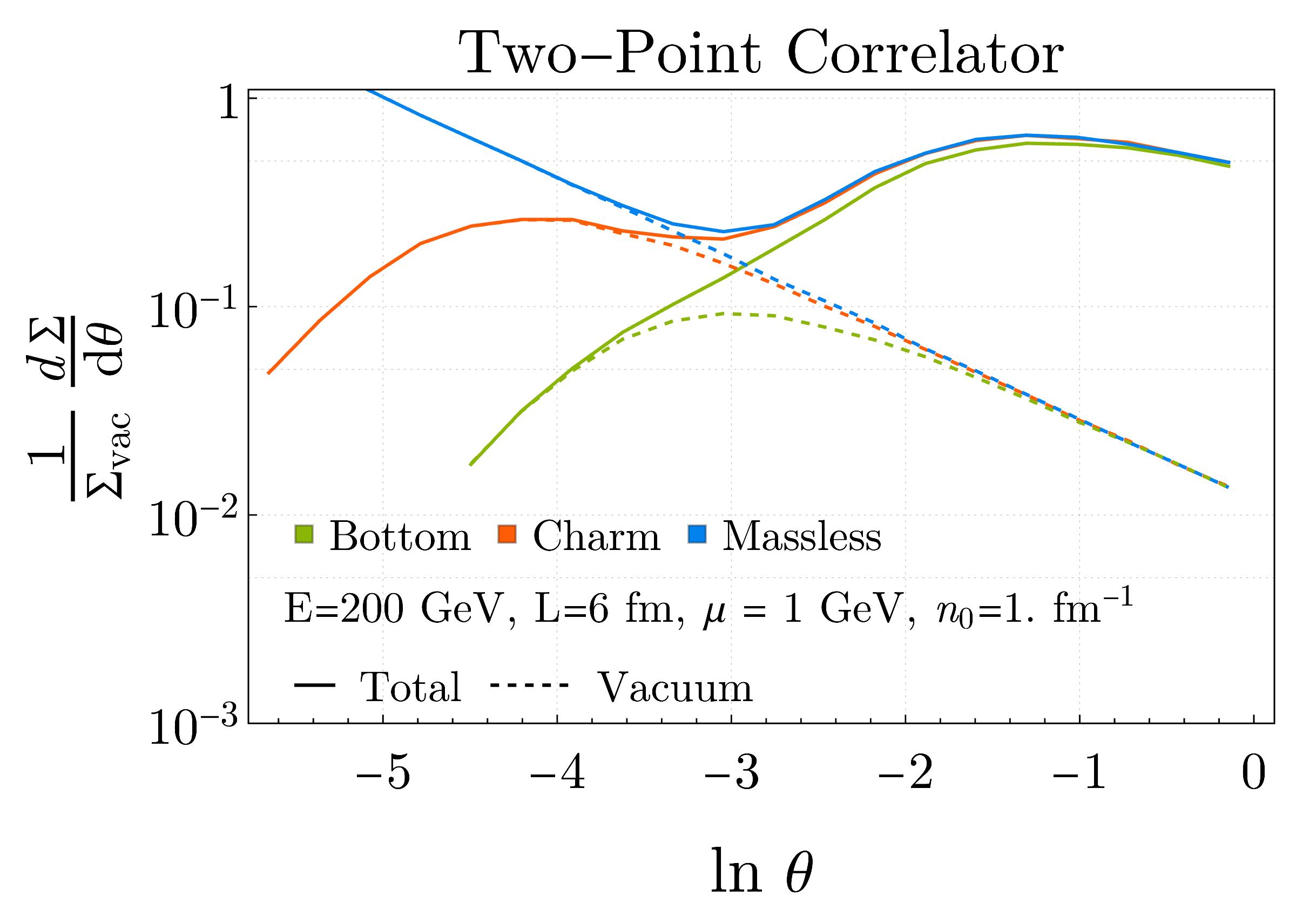} \\
\includegraphics[width=0.45\textwidth]{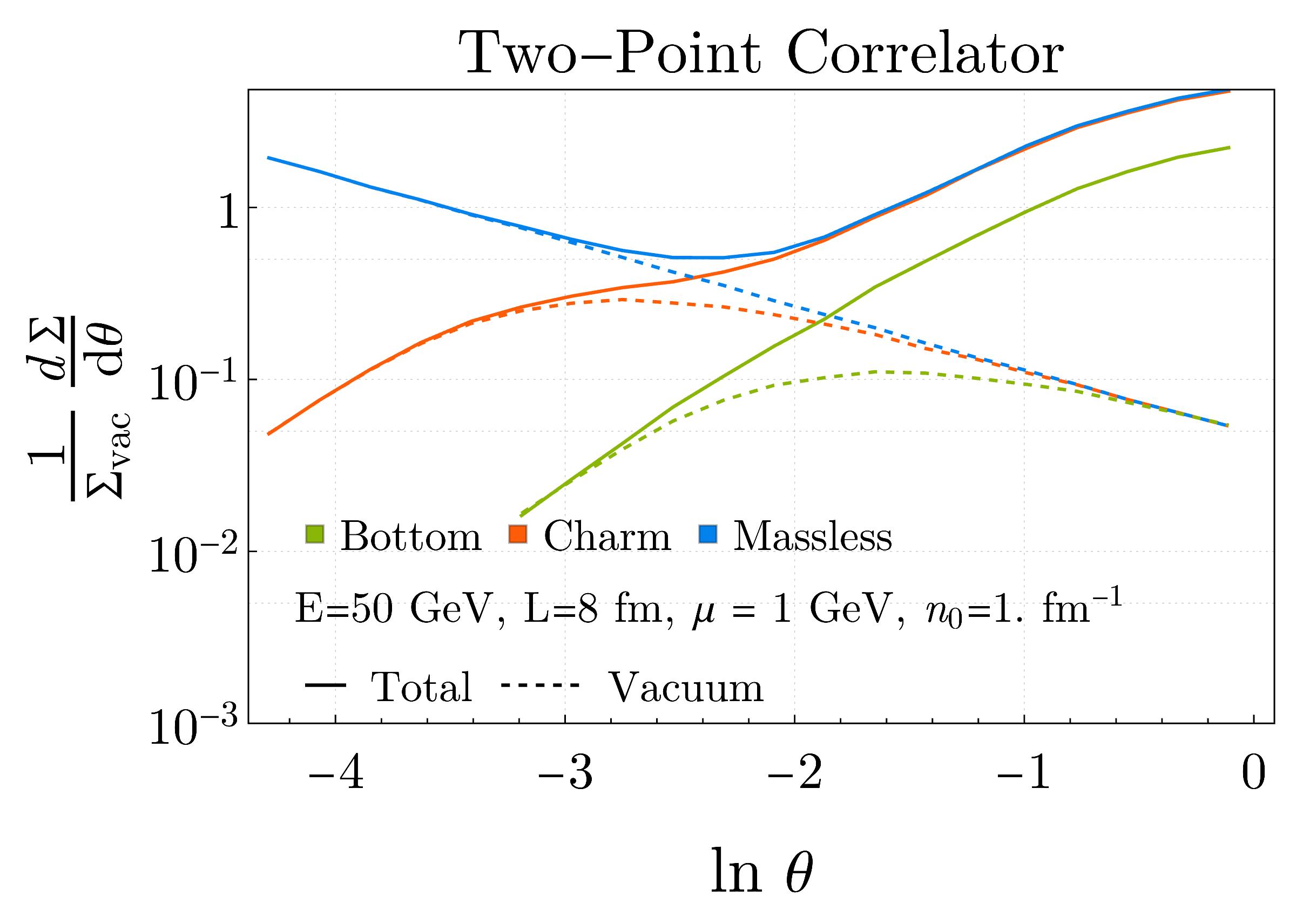} 
    \caption{EEC of a light-quark  (blue), $c$-quark (orange), and $b$-quark (green) jet in p-p (dashed) and heavy-ion (solid) collisions. Different panels correspond to different jet energies and medium parameters.
    All curves are normalised by the integrated vacuum result $\Sigma_{\rm vac}$.}
    \label{fig:spectra}
\end{figure}
%%%%%%%%%%%%%%%%%%%%%%%%%%%%%%%%%%%%%%%%%%%%%%%%%%%%%%%%%%%%%%%%%%%%%%

\emph{Numerical Results.}---In \Fig{fig:spectra} we compare the EEC for bottom, charm, and light quark initiated jets in heavy-ion and p-p collisions for different jet energies and medium parameters. Going from top to bottom, the ratio between the onset and dead-cone angles decreases. In the top panel there is a clear separation between the dead-cone suppression and the medium enhancement regions for both $b$ and $c$ heavy-ion jets. This is reflected in their medium-modified correlator as large changes in the slope of the power-law massless vacuum behavior (blue dashed). Since the two angular scales are well separated, the medium-modified EEC for heavy-quark jets can be seen as a combination of the enhancement at large angles obtained for the massless QGP case in \cite{Andres:2022ovj,Andres:2023xwr} and the depletion at small angles due to the dead-cone seen in the massive vacuum results in \cite{Craft:2022kdo}. In the middle panel, the separation between the two regions disappears for the $b$-jet, but not for the $c$-jet, as its dead-cone angle is smaller. As a result, the medium-modified EEC of the $c$-jet in this panel exhibits the same behavior as in the top panel, while for the $b$-jet we can clearly observe how medium-induced radiation starts to fill the dead-cone region. The detailed behavior of the bottom EEC in this regime, where both scales are relevant, cannot be reduced to either of the previously studied limiting cases. This finding highlights the intricate interplay between the two competing scales in the energy correlators and represents a novel result obtained from our calculation. Finally, in the lower panel where the separation between regions disappear for the $c$-jet too, we observe that medium-induced radiation has completely filled some of the angular extent of the dead-cone for the $b$-jet, and has started to fill the dead-cone for the $c$-jet. These distinct outcomes among the three parameter sets show the remarkable ability of energy correlators to resolve the ``filling of the dead-cone" \cite{Armesto:2003jh}, beautifully illustrating their potential in unveiling different scales within the QGP.

A particularly clean probe for studying the radiation off heavy-flavor jets in heavy-ion collisions is the ratio between the EECs of jets originating from quarks with different masses. This ratio is especially insightful when comparing $c$- and $b$-jets since, in both cases, the initiating quark can be traced through the QGP. This approach offers additional advantages from both the experimental and theoretical perspectives, as many experimental systematic uncertainties are expected to cancel out in the ratio, and the difference in mass between both quarks provides a well-defined perturbative window for studying medium modifications. 

%%%%%%%%%%%%%%%%%%%%%%%%%%%%%%%%%%%%%%%%%%%%%%%%%%
\begin{figure}[!]
\includegraphics[width=0.4\textwidth]{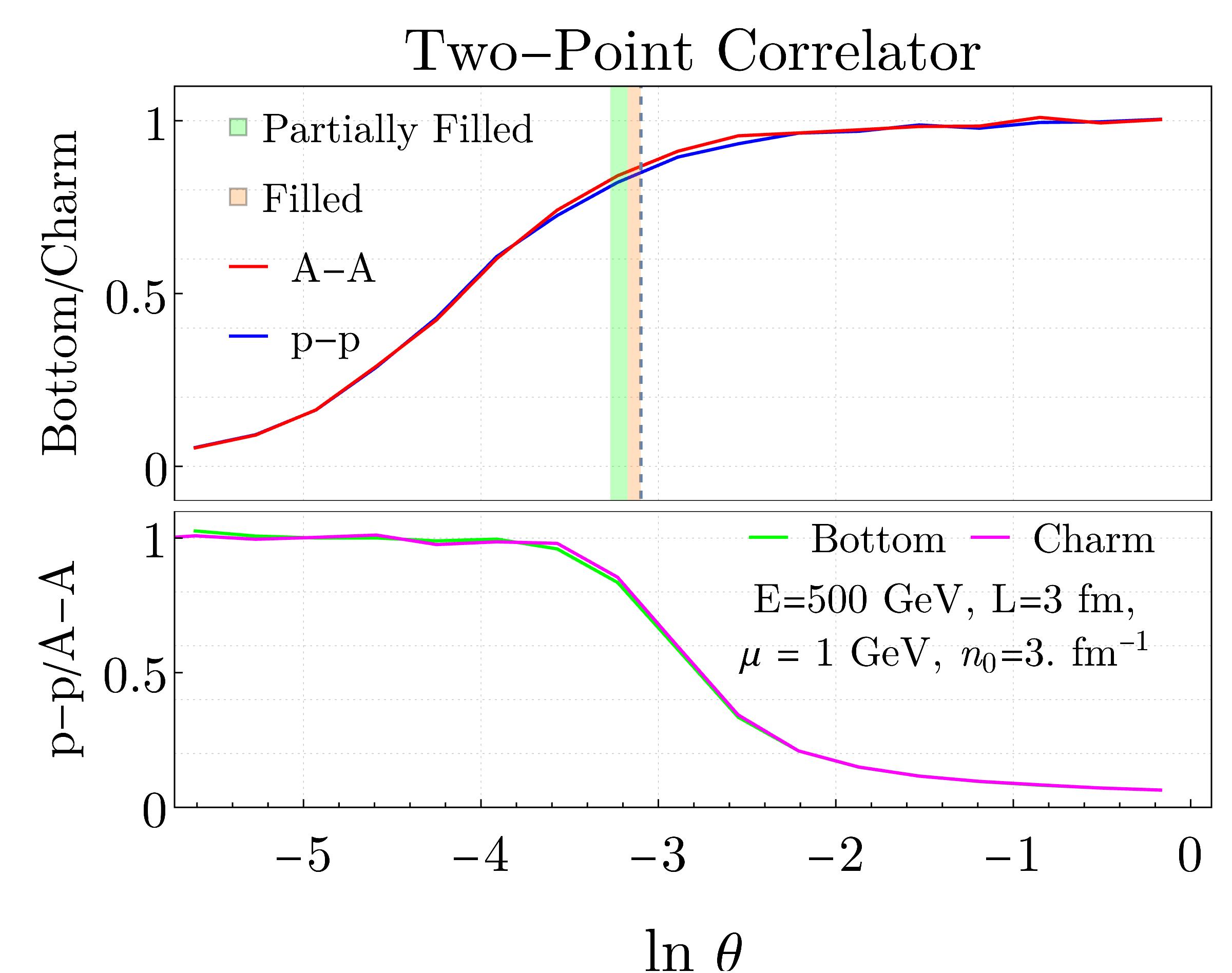}
\includegraphics[width=0.4\textwidth]{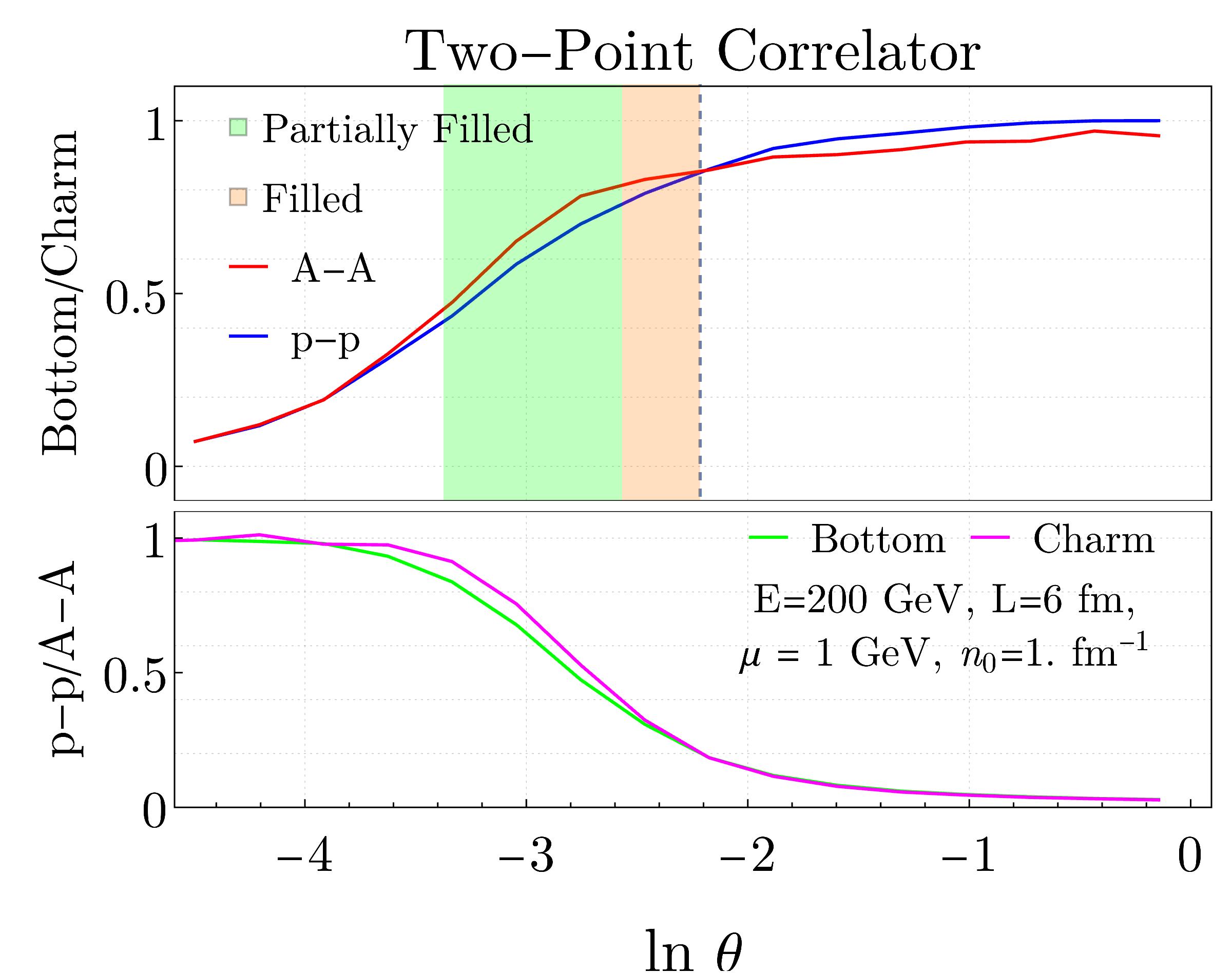}\\
\includegraphics[width=0.4\textwidth]{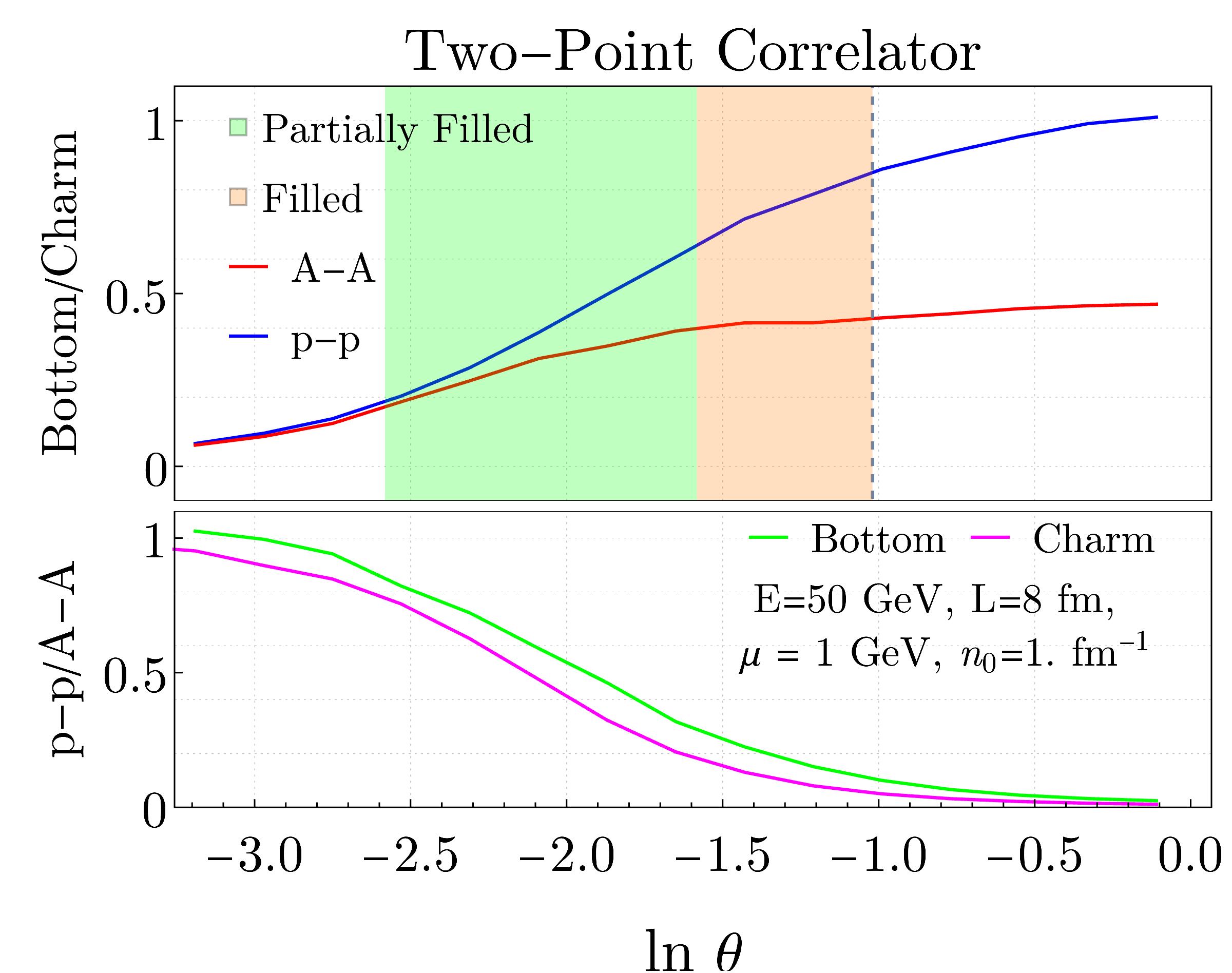}
\caption{Top half of each panel: ratio of the EEC of a $b$-jet w.r.t a $c$-jet in p-p (blue) and heavy-ion (red) collisions. Bottom half of each panel:  ratio of the EEC of $b$-jet ($c$-jet) in p-p w.r.t heavy-ion collisions in green (pink). The colored bands are meant to guide the eye towards the region where the dead-cone of the heavy-ion $b$-jet is filled, as explained in the text. }
\label{fig:zombie}
\end{figure}

In \Fig{fig:zombie} we show in the top half of each panel the ratio of the EEC of a bottom-quark jet with respect to a charm-quark jet, both in p-p and heavy-ion collisions. In the bottom halves, we present the ratio of the EEC of bottom- (charm-) quark jets in p-p with respect to heavy-ion collisions in green (pink). The parameters used in each panel agree with those of the corresponding panel in \Fig{fig:spectra}. Since we are interested in the overlap between the dead-cone depletion and the medium enhancement regions, we have introduced colored bands in the top halves of the panels to guide the eye towards this interplay for the $b$-jet. The dashed line, located at approximately $4m_b/E$, indicates the maximum angle at which the dead-cone effect is visible, while  the left boundary of the green region is a proxy for the onset of the medium enhancement, set to the angle where the A-A and p-p EECs for $b$-jets deviate by $15\%$ (where the green line in the bottom panels drops to 0.85). In the top panel, since the dead-cone and medium-induced regimes are distinctly separated for both $c$- and $b$-jets, the overlap region is practically non-existent and their radiation pattern within the dead-cone remains unmodified by the medium. Similarly, for this jet energy, the medium-induced radiation behaves as if the initial parton was massless, yielding an EEC $b/c$ ratio in A-A identical to the p-p result in the whole angular domain. Moving to the middle panel, the overlap region increases and medium-induced radiation begins to fill the dead-cone for the $b$-jet. This phenomenon is evident in the $b/c$ ratio, as it transitions from the low-angle to the green region, where it starts to deviate from the vacuum result due to medium-induced radiation. Although there is still a depletion of radiation due to the quark mass in this regime, the shape of the $b/c$ ratio is modified compared to the vacuum scenario. At sufficiently large angles, the A-A $b/c$ ratio becomes close to flat, reaching a saturated value outside the dead-cone. We highlight in red the part of this regime occurring within the dead-cone, representing the portion of the dead-cone regime that has been replenished by the medium-induced radiation. The boundary between the red and green regions is given by the angle where the A-A $b/c$ ratio reaches $85\%$ of the saturated value. The fact that its saturated value is less than 1 is indicative of the dependence of medium-induced radiation on the quark mass. In the lowest panel, we observe a similar behavior to the middle one, but with a larger fraction of the dead-cone filled with medium-induced radiation due to its lower value of $\theta_{\rm on}/\dead$. 
%%%%%%%%%%%%%%%%%%%%%%%%%%%%%%%%%%%%%%%%%%%%%%%%%%%%%%%%%%%%%%%%%%%%%%

%%%%%%%%%%%%%%%%%%%%%%%%%%%%%%%%%%%%%%%%%%%%%%%%%%%%%%%%%%%%%%%%%%%%%
\begin{figure}[ht]
\includegraphics[width=0.4\textwidth]{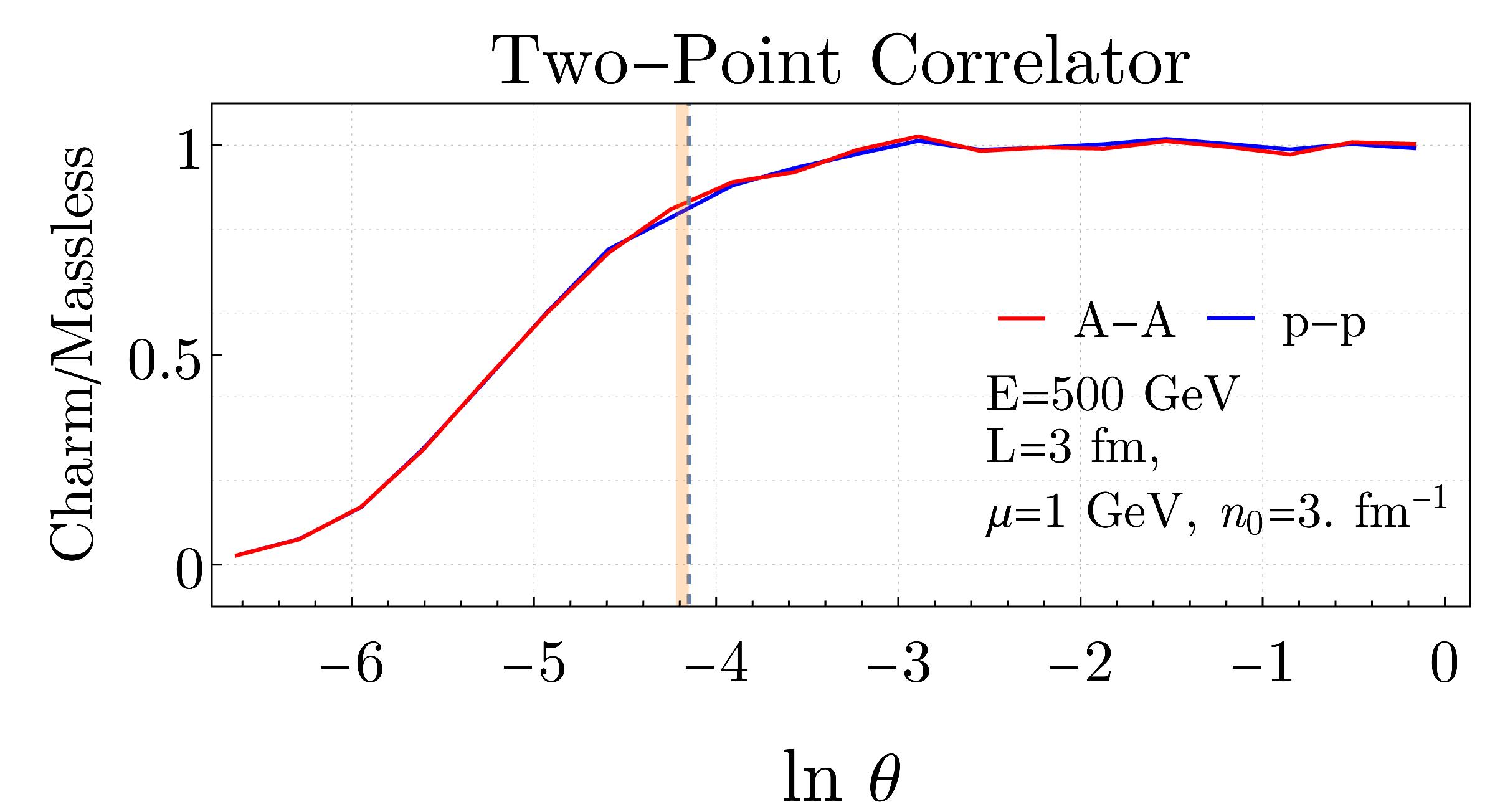}\\
\includegraphics[width=0.4\textwidth]{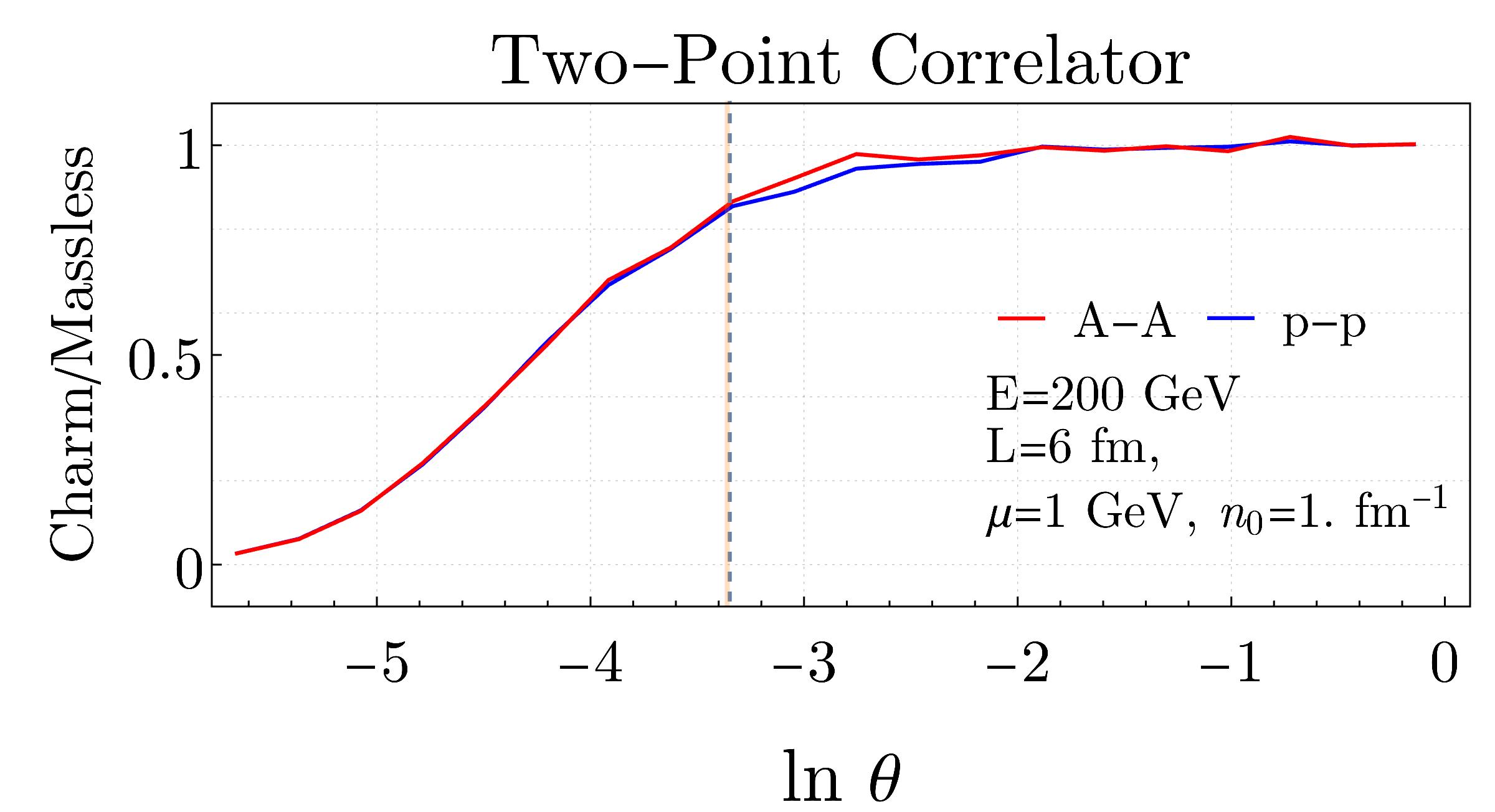}\\
\includegraphics[width=0.4\textwidth]{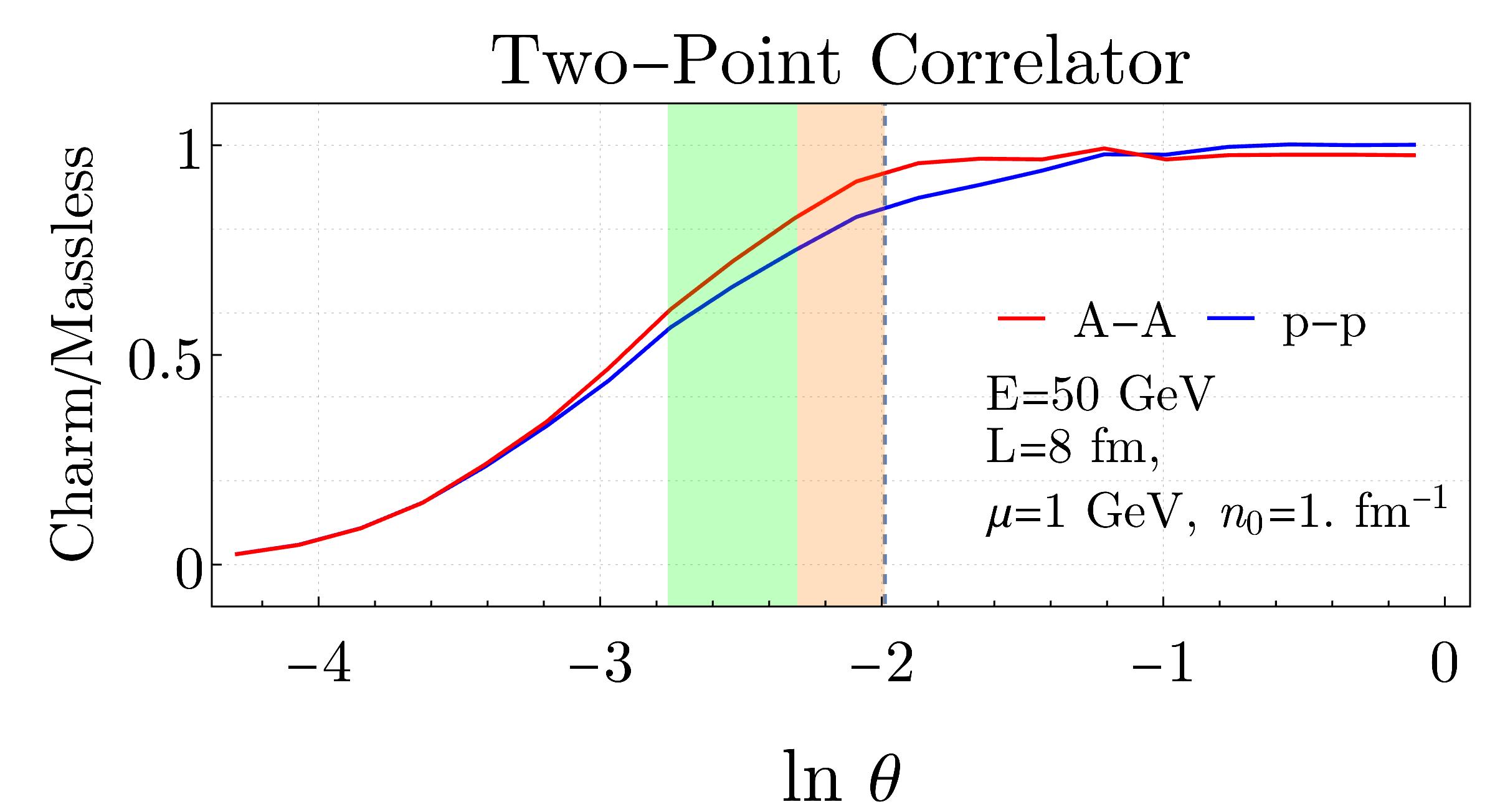}
\caption{Ratio of the EEC correlator of a $c$-jet w.r.t light-quark jet in p-p (blue) and heavy-ion (red) collisions. Each panel corresponds to different jet energies and medium parameters. The colored bands are meant to guide the eye towards the region where the dead-cone of the heavy-ion $c$-jet is filled by medium-induced radiation. They are defined as explained in the text for \Fig{fig:zombie} replacing bottom with charm and charm with massless.}
\label{fig:charmtomassless}
\end{figure}
%%%%%%%%%%%%%%%%%%%%%%%%%%%%%%%%%%%%%%%%%%%%%%%%%%
%%%%%%%%%%%%%%%%%%%
Although a large sample of heavy-ion $b$-jets are expected to be available in the near future, since $b$-jets are rarer, a similar analysis can also be conducted using the ratio between $c$-jets and massless jets. Indeed, ALICE used charm-to-massless ratios of the splitting angle distribution for the measurement of the dead-cone in p-p \cite{ALICE:2021aqk}. This approach will suffer some additional complexity due the $c$ dead-cone being found at smaller angles than the $b$ dead-cone and to the uncertainty in the $q/g$ fraction of massless jets. However, the leading order p-p baseline for this observable does not depend on the $q/g$ fraction \cite{Craft:2022kdo}, and the increase of the $q$ fraction due to energy loss in heavy-ions is expected to lead to only mild logarithmic modifications on the medium-modified EEC ~\cite{Andres:2022ovj,Andres:2023xwr}.

As this will be the first experimentally viable measurement, we present the results of our calculations for this case in \Fig{fig:charmtomassless}. We observe similar features as in \Fig{fig:zombie}, but due to the smaller dead-cone for $c$-jets, the effect of the medium is comparatively smaller.  To observe the filling of the dead-cone, it will be necessary to employ lower energy jets, as shown in the lowest panel of \Fig{fig:charmtomassless}.

%%%%%%%%%%%%%%%%%%%%%%%%%%%%%%%%%%%%%%%%%%%%%%%%%%%%%%%%%%%%%%%%%%%%%%%%%%%%
\emph{Conclusions.}--- In this \emph{Letter}, we propose a new observable to search for dead-cone effects in heavy-flavor jets within heavy-ion collisions. Our proposed observable relies on energy correlators, which we consider to be highly suitable for this task owing to their distinct ability to separate the perturbative and non-perturbative regimes \cite{Komiske:2022enw}, as well as their reduced sensitivity to soft physics. Specifically, we have analyzed the two-point energy correlator of a heavy-flavor jet in a simplified model and showed its sensitivity to the filling of the dead-cone region by medium-induced emissions.  Notably, when the dead-cone depletion and medium-induced enhancement regions overlap, the correlator displays new structure which cannot be reduced to the sum of the two phenomena. This analysis shows the remarkable potential of energy correlators to disentangle different separated scales, representing the first example where energy correlators simultaneously resolve both an intrinsic scale, associated with the heavy quark, and an emergent scale  originating from the interactions with the medium.

Our present work uses some simplifying assumptions, including the use of a static brick of plasma, which will be relaxed in subsequent studies. More comprehensive analyses are required to quantitatively assess the impact of different effects on the correlator, such as its resilience to the underlying background or medium response. Despite these considerations, the results of this study are an important first step towards understanding energy correlators in a more realistic environment and should motivate future phenomenological studies.

With the recent measurements of energy correlators in p-p jets \cite{talk1,talk2,talk3}, and the forthcoming data on light- and heavy-flavor jets in heavy-ion collisions at RHIC and the LHC, we anticipate that energy correlators will play a crucial role in unraveling their radiation patterns, leading to new insights into the microscopic nature of the QGP.

%%%%%%%%%%%%%%%%%%%%%%%%%%%%%%%%%%%%%%%%%%%%%%%%%%%%%%%%%%%%%%%%%%%%%%%%%%%%
\emph{Acknowledgements.}---We thank Raghav Kunnawalkam Elayavalli, Leticia Cunqueiro, Mateusz Ploskon, Wenqing Fan, Laura Havener, Ananya Rai, and Xin-Nian Wang for useful discussions. 
This work is supported by  European Research Council project ERC-2018-ADG-835105 YoctoLHC; by Maria de Maetzu excellence program under project CEX2020-001035-M; by Spanish Research State Agency under project PID2020-119632GB-I00; by European Union ERDF; by Xunta de Galicia (CIGUS Network of Research Centers). This work is supported in part by the GLUODYNAMICS project funded by the ``P2IO LabEx (ANR-10-LABX-0038)'' in the framework ``Investissements d'Avenir'' (ANR-11-IDEX-0003-01) managed by the Agence Nationale de la Recherche (ANR), France. I.M. is supported by start-up funds from Yale University. C.A. has received funding from the European Union’s Horizon 2020 research and innovation program under the Marie Sklodowska-Curie grant agreement No 893021 (JQ4LHC).
%%%%%%%%%%%%%%%%%%%%%%%%%%%%%%%%%%%%%%%%%%%%%%%%%%%%%%%%%%%%%%%%%%%%%%%%%%%%

\appendix

\section{Appendix A}

%%%%%%%%%%%%%%%%%%%%%%%%%%%%%%%%%%%%%%%%%%%%%%%%%%%%%%%%%%%%%%%%%%%%%%
\begin{figure}[ht]
\includegraphics[width=0.45\textwidth]{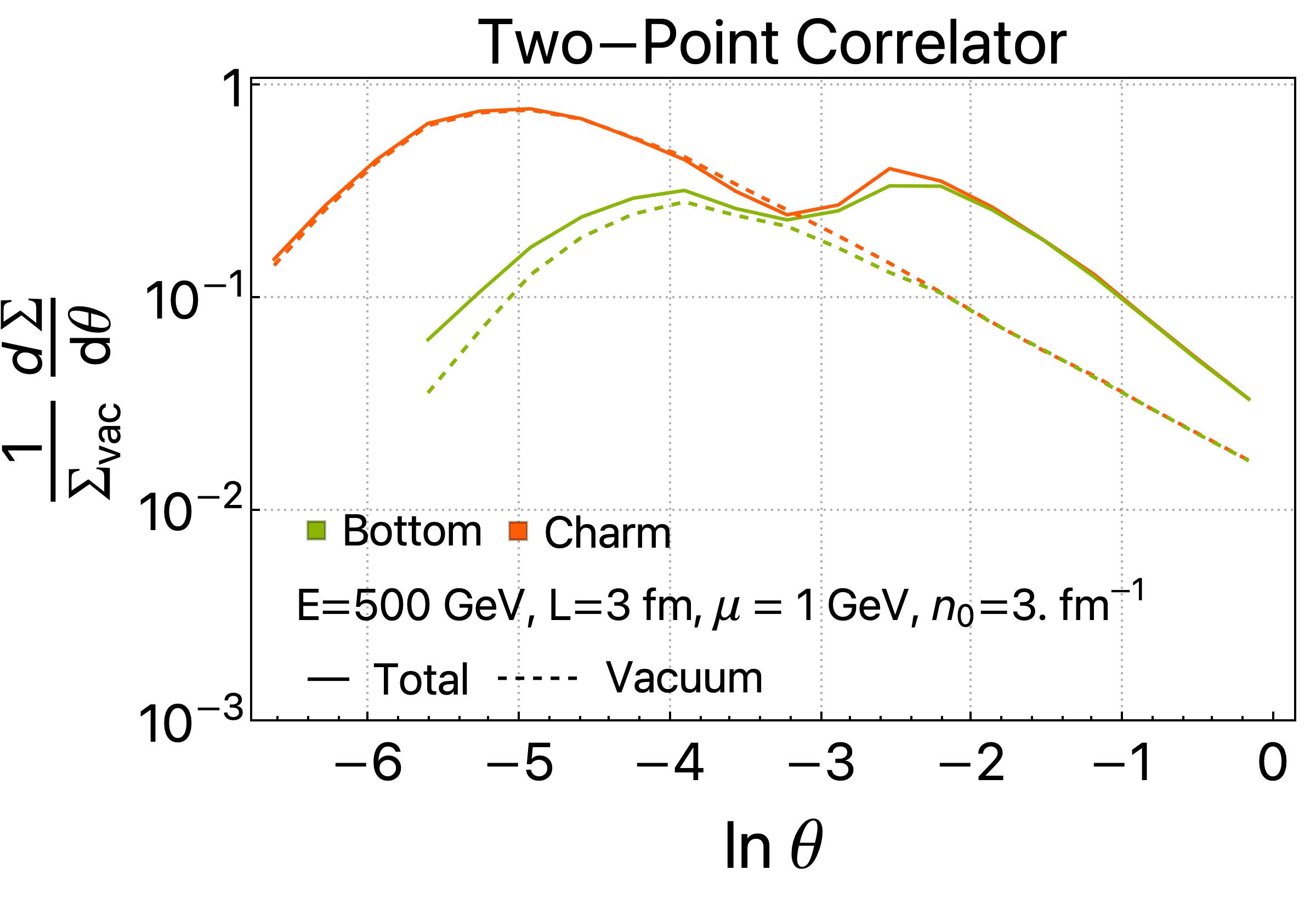}\\
\includegraphics[width=0.45\textwidth]{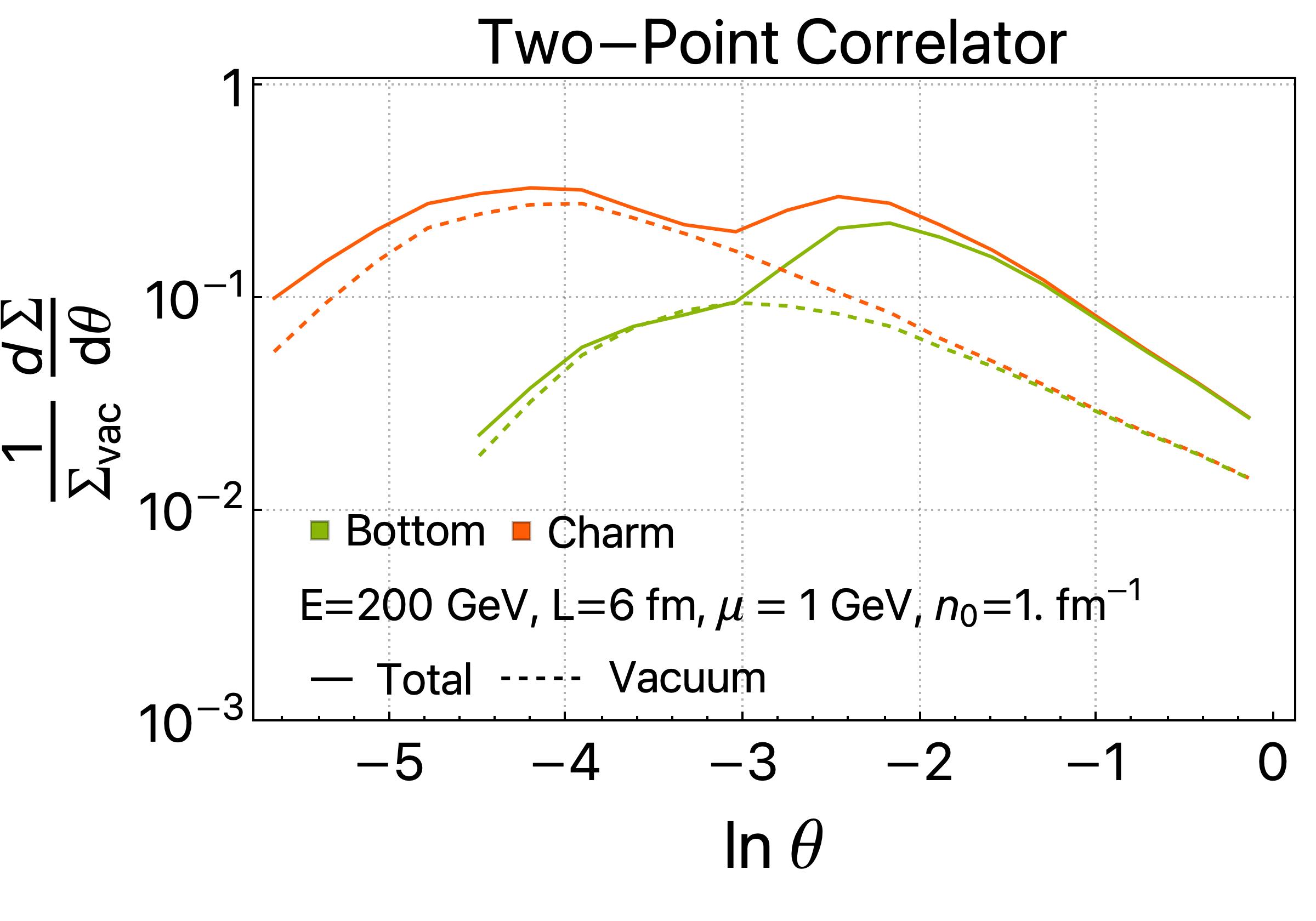} \\
\includegraphics[width=0.45\textwidth]{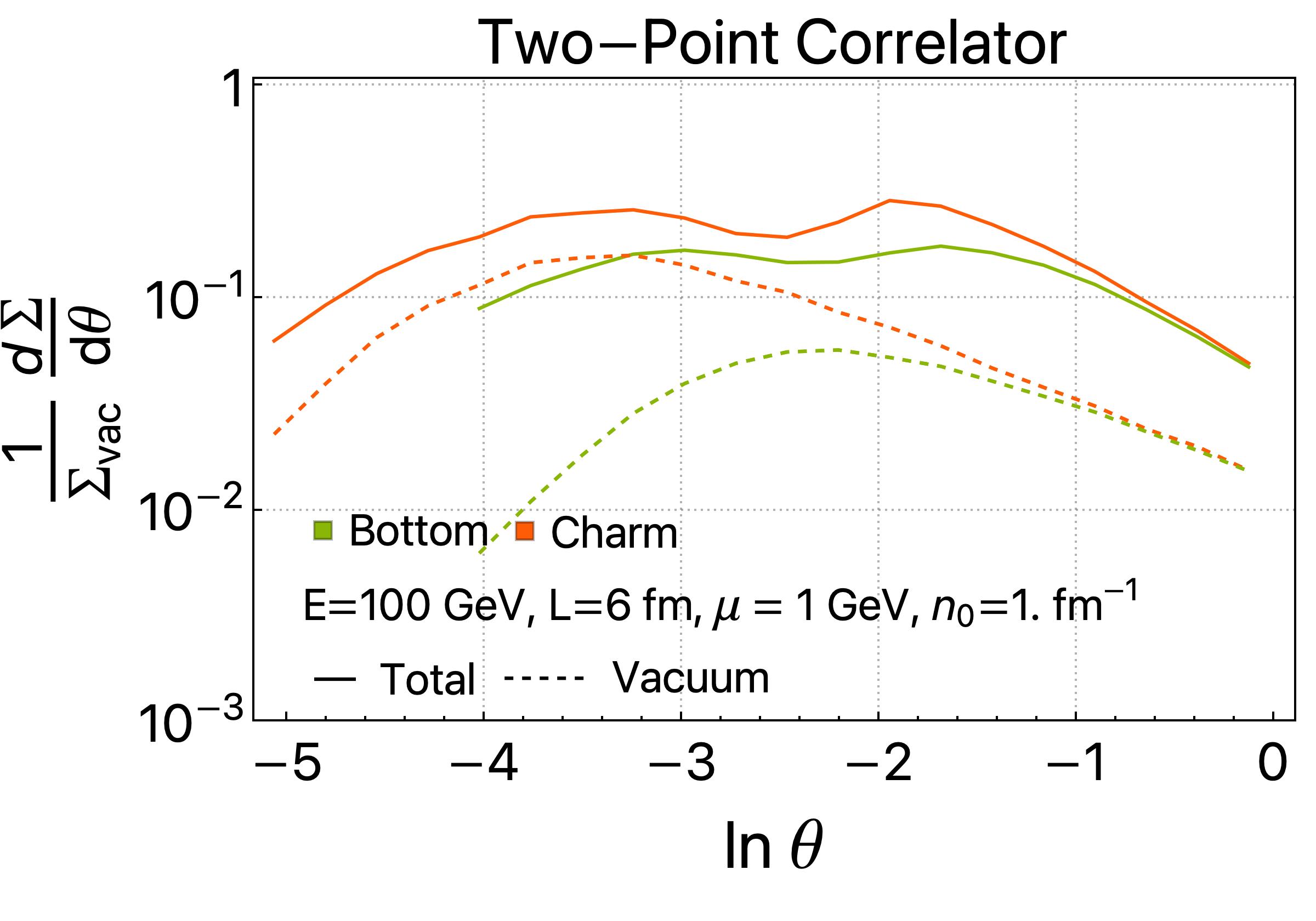} 
    \caption{EEC of a $c$-quark (orange), and $b$-quark (green) jet in p-p (dashed) and heavy-ion (solid) collisions. The curves were obtained using the single scattering approximation for the medium-induced splitting. Different panels correspond to different jet energies and medium parameters.
    All curves are normalised by the integrated vacuum result $\Sigma_{\rm vac}$.}
    \label{fig:singlescattering}
\end{figure}
%%%%%%%%%%%%%%%%%%%%%%%%%%%%%%%%%%%%%%%%%%%%%%%%%%%%%%%%%%%%%%%%%%%%%%

In this appendix we present an alternative computation of the curves from Fig.~\ref{fig:spectra} using the single scattering approximation for $F_{\rm med}$ with a massive quark jet. An explicit expression for $F_{\rm med}$ can be easily derived from Eq.~(59) of \cite{Sievert:2018imd}.
%\footnote{More precisely, Eq.~(59) gives $\frac{\td \hat\sigma^{\rm vac}_{qg}}{\td \theta \td z } F_{\rm med}$ up to a Jacobian.} 
The mass dependence in this expression is substantially more complicated than in the semi-hard approximation of the multiple scattering BDMPS-Z formalism. Of particular note, in the limit that the angle becomes very small, $F_{\rm med}$ saturates to a non-zero function of the medium parameters, whilst the semi-hard approximation vanishes due to the coherent effects among multiple scatterings. Additionally, in the limit where the jet energy becomes small, the single scattering approximation becomes badly behaved since the magnitude of $F_{\rm med}$ grows at all points in the emission phase-space --- leading to negative cross-sections. Consequently, we set the lowest jet energy considered within this approach to 100~GeV, instead of the 50~GeV used in the semi-hard multiple scattering formalism. The results for the single scattering approximation are shown in Fig.~\ref{fig:singlescattering}. Despite the increased complexity, and the different limiting behaviour, we find qualitatively consistent results between the semi-hard and single scattering approximations. In both approaches, when $\theta_{\rm on} \gg m_{Q}/E$ we see a clear separation between regimes. Specifically, at angles $\sim \theta_{\rm on}$ the correlator spectrum is sensitive to the medium rather than the mass, while at angles $\sim m_{Q}/E$ it is sensitive to the dead-cone effect and not the medium. In contrast, when $\theta_{\rm on} \sim m_{Q}/E$, the medium's imprint on the spectrum overlaps with the dead-cone, which can  be filled. We note that the exact nature of the filling does appears to be model dependent.

%%%%%%%%%%%%%%%%%%%%%%%%%%%%%%%%%%%%%%%%%%%%%%%%%%%%%%%%%%%%%%%%%%%%%%%%%%%%
\bibliography{qgp_EEC_ref.bib}{}

\begin{thebibliography}{98}%
\makeatletter
\providecommand \@ifxundefined [1]{%
 \@ifx{#1\undefined}
}%
\providecommand \@ifnum [1]{%
 \ifnum #1\expandafter \@firstoftwo
 \else \expandafter \@secondoftwo
 \fi
}%
\providecommand \@ifx [1]{%
 \ifx #1\expandafter \@firstoftwo
 \else \expandafter \@secondoftwo
 \fi
}%
\providecommand \natexlab [1]{#1}%
\providecommand \enquote  [1]{``#1''}%
\providecommand \bibnamefont  [1]{#1}%
\providecommand \bibfnamefont [1]{#1}%
\providecommand \citenamefont [1]{#1}%
\providecommand \href@noop [0]{\@secondoftwo}%
\providecommand \href [0]{\begingroup \@sanitize@url \@href}%
\providecommand \@href[1]{\@@startlink{#1}\@@href}%
\providecommand \@@href[1]{\endgroup#1\@@endlink}%
\providecommand \@sanitize@url [0]{\catcode `\\12\catcode `\$12\catcode
  `\&12\catcode `\#12\catcode `\^12\catcode `\_12\catcode `\%12\relax}%
\providecommand \@@startlink[1]{}%
\providecommand \@@endlink[0]{}%
\providecommand \url  [0]{\begingroup\@sanitize@url \@url }%
\providecommand \@url [1]{\endgroup\@href {#1}{\urlprefix }}%
\providecommand \urlprefix  [0]{URL }%
\providecommand \Eprint [0]{\href }%
\providecommand \doibase [0]{http://dx.doi.org/}%
\providecommand \selectlanguage [0]{\@gobble}%
\providecommand \bibinfo  [0]{\@secondoftwo}%
\providecommand \bibfield  [0]{\@secondoftwo}%
\providecommand \translation [1]{[#1]}%
\providecommand \BibitemOpen [0]{}%
\providecommand \bibitemStop [0]{}%
\providecommand \bibitemNoStop [0]{.\EOS\space}%
\providecommand \EOS [0]{\spacefactor3000\relax}%
\providecommand \BibitemShut  [1]{\csname bibitem#1\endcsname}%
\let\auto@bib@innerbib\@empty
%</preamble>
\bibitem [{\citenamefont {Gyulassy}\ and\ \citenamefont
  {McLerran}(2005)}]{Gyulassy:2004zy}%
  \BibitemOpen
  \bibfield  {author} {\bibinfo {author} {\bibfnamefont {M.}~\bibnamefont
  {Gyulassy}}\ and\ \bibinfo {author} {\bibfnamefont {L.}~\bibnamefont
  {McLerran}},\ }\href {\doibase 10.1016/j.nuclphysa.2004.10.034} {\bibfield
  {journal} {\bibinfo  {journal} {Nucl. Phys. A}\ }\textbf {\bibinfo {volume}
  {750}},\ \bibinfo {pages} {30} (\bibinfo {year} {2005})},\ \Eprint
  {http://arxiv.org/abs/nucl-th/0405013} {arXiv:nucl-th/0405013} \BibitemShut
  {NoStop}%
\bibitem [{\citenamefont {Back}\ \emph {et~al.}(2005)\citenamefont {Back} \emph
  {et~al.}}]{PHOBOS:2004zne}%
  \BibitemOpen
  \bibfield  {author} {\bibinfo {author} {\bibfnamefont {B.~B.}\ \bibnamefont
  {Back}} \emph {et~al.} (\bibinfo {collaboration} {PHOBOS}),\ }\href {\doibase
  10.1016/j.nuclphysa.2005.03.084} {\bibfield  {journal} {\bibinfo  {journal}
  {Nucl. Phys. A}\ }\textbf {\bibinfo {volume} {757}},\ \bibinfo {pages} {28}
  (\bibinfo {year} {2005})},\ \Eprint {http://arxiv.org/abs/nucl-ex/0410022}
  {arXiv:nucl-ex/0410022} \BibitemShut {NoStop}%
\bibitem [{\citenamefont {Muller}\ and\ \citenamefont
  {Nagle}(2006)}]{Muller:2006ee}%
  \BibitemOpen
  \bibfield  {author} {\bibinfo {author} {\bibfnamefont {B.}~\bibnamefont
  {Muller}}\ and\ \bibinfo {author} {\bibfnamefont {J.~L.}\ \bibnamefont
  {Nagle}},\ }\href {\doibase 10.1146/annurev.nucl.56.080805.140556} {\bibfield
   {journal} {\bibinfo  {journal} {Ann. Rev. Nucl. Part. Sci.}\ }\textbf
  {\bibinfo {volume} {56}},\ \bibinfo {pages} {93} (\bibinfo {year} {2006})},\
  \Eprint {http://arxiv.org/abs/nucl-th/0602029} {arXiv:nucl-th/0602029}
  \BibitemShut {NoStop}%
\bibitem [{\citenamefont {Muller}\ \emph {et~al.}(2012)\citenamefont {Muller},
  \citenamefont {Schukraft},\ and\ \citenamefont {Wyslouch}}]{Muller:2012zq}%
  \BibitemOpen
  \bibfield  {author} {\bibinfo {author} {\bibfnamefont {B.}~\bibnamefont
  {Muller}}, \bibinfo {author} {\bibfnamefont {J.}~\bibnamefont {Schukraft}}, \
  and\ \bibinfo {author} {\bibfnamefont {B.}~\bibnamefont {Wyslouch}},\ }\href
  {\doibase 10.1146/annurev-nucl-102711-094910} {\bibfield  {journal} {\bibinfo
   {journal} {Ann. Rev. Nucl. Part. Sci.}\ }\textbf {\bibinfo {volume} {62}},\
  \bibinfo {pages} {361} (\bibinfo {year} {2012})},\ \Eprint
  {http://arxiv.org/abs/1202.3233} {arXiv:1202.3233 [hep-ex]} \BibitemShut
  {NoStop}%
\bibitem [{\citenamefont {Busza}\ \emph {et~al.}(2018)\citenamefont {Busza},
  \citenamefont {Rajagopal},\ and\ \citenamefont {van~der
  Schee}}]{Busza:2018rrf}%
  \BibitemOpen
  \bibfield  {author} {\bibinfo {author} {\bibfnamefont {W.}~\bibnamefont
  {Busza}}, \bibinfo {author} {\bibfnamefont {K.}~\bibnamefont {Rajagopal}}, \
  and\ \bibinfo {author} {\bibfnamefont {W.}~\bibnamefont {van~der Schee}},\
  }\href {\doibase 10.1146/annurev-nucl-101917-020852} {\bibfield  {journal}
  {\bibinfo  {journal} {Ann. Rev. Nucl. Part. Sci.}\ }\textbf {\bibinfo
  {volume} {68}},\ \bibinfo {pages} {339} (\bibinfo {year} {2018})},\ \Eprint
  {http://arxiv.org/abs/1802.04801} {arXiv:1802.04801 [hep-ph]} \BibitemShut
  {NoStop}%
\bibitem [{\citenamefont {Dexheimer}\ \emph {et~al.}(2021)\citenamefont
  {Dexheimer}, \citenamefont {Noronha}, \citenamefont {Noronha-Hostler},
  \citenamefont {Ratti},\ and\ \citenamefont {Yunes}}]{Dexheimer:2020zzs}%
  \BibitemOpen
  \bibfield  {author} {\bibinfo {author} {\bibfnamefont {V.}~\bibnamefont
  {Dexheimer}}, \bibinfo {author} {\bibfnamefont {J.}~\bibnamefont {Noronha}},
  \bibinfo {author} {\bibfnamefont {J.}~\bibnamefont {Noronha-Hostler}},
  \bibinfo {author} {\bibfnamefont {C.}~\bibnamefont {Ratti}}, \ and\ \bibinfo
  {author} {\bibfnamefont {N.}~\bibnamefont {Yunes}},\ }\href {\doibase
  10.1088/1361-6471/abe104} {\bibfield  {journal} {\bibinfo  {journal} {J.
  Phys. G}\ }\textbf {\bibinfo {volume} {48}},\ \bibinfo {pages} {073001}
  (\bibinfo {year} {2021})},\ \Eprint {http://arxiv.org/abs/2010.08834}
  {arXiv:2010.08834 [nucl-th]} \BibitemShut {NoStop}%
\bibitem [{\citenamefont {Arslandok}\ \emph {et~al.}(2023)\citenamefont
  {Arslandok} \emph {et~al.}}]{Arslandok:2023utm}%
  \BibitemOpen
  \bibfield  {author} {\bibinfo {author} {\bibfnamefont {M.}~\bibnamefont
  {Arslandok}} \emph {et~al.},\ }\href@noop {} {\  (\bibinfo {year} {2023})},\
  \Eprint {http://arxiv.org/abs/2303.17254} {arXiv:2303.17254 [nucl-ex]}
  \BibitemShut {NoStop}%
\bibitem [{\citenamefont {Larkoski}\ \emph {et~al.}(2020)\citenamefont
  {Larkoski}, \citenamefont {Moult},\ and\ \citenamefont
  {Nachman}}]{Larkoski:2017jix}%
  \BibitemOpen
  \bibfield  {author} {\bibinfo {author} {\bibfnamefont {A.~J.}\ \bibnamefont
  {Larkoski}}, \bibinfo {author} {\bibfnamefont {I.}~\bibnamefont {Moult}}, \
  and\ \bibinfo {author} {\bibfnamefont {B.}~\bibnamefont {Nachman}},\ }\href
  {\doibase 10.1016/j.physrep.2019.11.001} {\bibfield  {journal} {\bibinfo
  {journal} {Phys. Rept.}\ }\textbf {\bibinfo {volume} {841}},\ \bibinfo
  {pages} {1} (\bibinfo {year} {2020})},\ \Eprint
  {http://arxiv.org/abs/1709.04464} {arXiv:1709.04464 [hep-ph]} \BibitemShut
  {NoStop}%
\bibitem [{\citenamefont {Kogler}\ \emph {et~al.}(2019)\citenamefont {Kogler}
  \emph {et~al.}}]{Kogler:2018hem}%
  \BibitemOpen
  \bibfield  {author} {\bibinfo {author} {\bibfnamefont {R.}~\bibnamefont
  {Kogler}} \emph {et~al.},\ }\href {\doibase 10.1103/RevModPhys.91.045003}
  {\bibfield  {journal} {\bibinfo  {journal} {Rev. Mod. Phys.}\ }\textbf
  {\bibinfo {volume} {91}},\ \bibinfo {pages} {045003} (\bibinfo {year}
  {2019})},\ \Eprint {http://arxiv.org/abs/1803.06991} {arXiv:1803.06991
  [hep-ex]} \BibitemShut {NoStop}%
\bibitem [{\citenamefont {Connors}\ \emph {et~al.}(2018)\citenamefont
  {Connors}, \citenamefont {Nattrass}, \citenamefont {Reed},\ and\
  \citenamefont {Salur}}]{Connors:2017ptx}%
  \BibitemOpen
  \bibfield  {author} {\bibinfo {author} {\bibfnamefont {M.}~\bibnamefont
  {Connors}}, \bibinfo {author} {\bibfnamefont {C.}~\bibnamefont {Nattrass}},
  \bibinfo {author} {\bibfnamefont {R.}~\bibnamefont {Reed}}, \ and\ \bibinfo
  {author} {\bibfnamefont {S.}~\bibnamefont {Salur}},\ }\href {\doibase
  10.1103/RevModPhys.90.025005} {\bibfield  {journal} {\bibinfo  {journal}
  {Rev. Mod. Phys.}\ }\textbf {\bibinfo {volume} {90}},\ \bibinfo {pages}
  {025005} (\bibinfo {year} {2018})},\ \Eprint
  {http://arxiv.org/abs/1705.01974} {arXiv:1705.01974 [nucl-ex]} \BibitemShut
  {NoStop}%
\bibitem [{\citenamefont {Cunqueiro}\ and\ \citenamefont
  {Sickles}(2022)}]{Cunqueiro:2021wls}%
  \BibitemOpen
  \bibfield  {author} {\bibinfo {author} {\bibfnamefont {L.}~\bibnamefont
  {Cunqueiro}}\ and\ \bibinfo {author} {\bibfnamefont {A.~M.}\ \bibnamefont
  {Sickles}},\ }\href {\doibase 10.1016/j.ppnp.2022.103940} {\bibfield
  {journal} {\bibinfo  {journal} {Prog. Part. Nucl. Phys.}\ }\textbf {\bibinfo
  {volume} {124}},\ \bibinfo {pages} {103940} (\bibinfo {year} {2022})},\
  \Eprint {http://arxiv.org/abs/2110.14490} {arXiv:2110.14490 [nucl-ex]}
  \BibitemShut {NoStop}%
\bibitem [{\citenamefont {Djordjevic}\ and\ \citenamefont
  {Gyulassy}(2004)}]{Djordjevic:2003zk}%
  \BibitemOpen
  \bibfield  {author} {\bibinfo {author} {\bibfnamefont {M.}~\bibnamefont
  {Djordjevic}}\ and\ \bibinfo {author} {\bibfnamefont {M.}~\bibnamefont
  {Gyulassy}},\ }\href {\doibase 10.1016/j.nuclphysa.2003.12.020} {\bibfield
  {journal} {\bibinfo  {journal} {Nucl. Phys. A}\ }\textbf {\bibinfo {volume}
  {733}},\ \bibinfo {pages} {265} (\bibinfo {year} {2004})},\ \Eprint
  {http://arxiv.org/abs/nucl-th/0310076} {arXiv:nucl-th/0310076} \BibitemShut
  {NoStop}%
\bibitem [{\citenamefont {Djordjevic}\ \emph {et~al.}(2005)\citenamefont
  {Djordjevic}, \citenamefont {Gyulassy},\ and\ \citenamefont
  {Wicks}}]{Djordjevic:2004nq}%
  \BibitemOpen
  \bibfield  {author} {\bibinfo {author} {\bibfnamefont {M.}~\bibnamefont
  {Djordjevic}}, \bibinfo {author} {\bibfnamefont {M.}~\bibnamefont
  {Gyulassy}}, \ and\ \bibinfo {author} {\bibfnamefont {S.}~\bibnamefont
  {Wicks}},\ }\href {\doibase 10.1103/PhysRevLett.94.112301} {\bibfield
  {journal} {\bibinfo  {journal} {Phys. Rev. Lett.}\ }\textbf {\bibinfo
  {volume} {94}},\ \bibinfo {pages} {112301} (\bibinfo {year} {2005})},\
  \Eprint {http://arxiv.org/abs/hep-ph/0410372} {arXiv:hep-ph/0410372}
  \BibitemShut {NoStop}%
\bibitem [{\citenamefont {Armesto}\ \emph {et~al.}(2004)\citenamefont
  {Armesto}, \citenamefont {Salgado},\ and\ \citenamefont
  {Wiedemann}}]{Armesto:2003jh}%
  \BibitemOpen
  \bibfield  {author} {\bibinfo {author} {\bibfnamefont {N.}~\bibnamefont
  {Armesto}}, \bibinfo {author} {\bibfnamefont {C.~A.}\ \bibnamefont
  {Salgado}}, \ and\ \bibinfo {author} {\bibfnamefont {U.~A.}\ \bibnamefont
  {Wiedemann}},\ }\href {\doibase 10.1103/PhysRevD.69.114003} {\bibfield
  {journal} {\bibinfo  {journal} {Phys. Rev. D}\ }\textbf {\bibinfo {volume}
  {69}},\ \bibinfo {pages} {114003} (\bibinfo {year} {2004})},\ \Eprint
  {http://arxiv.org/abs/hep-ph/0312106} {arXiv:hep-ph/0312106} \BibitemShut
  {NoStop}%
\bibitem [{\citenamefont {Zhang}\ \emph {et~al.}(2004)\citenamefont {Zhang},
  \citenamefont {Wang},\ and\ \citenamefont {Wang}}]{Zhang:2003wk}%
  \BibitemOpen
  \bibfield  {author} {\bibinfo {author} {\bibfnamefont {B.-W.}\ \bibnamefont
  {Zhang}}, \bibinfo {author} {\bibfnamefont {E.}~\bibnamefont {Wang}}, \ and\
  \bibinfo {author} {\bibfnamefont {X.-N.}\ \bibnamefont {Wang}},\ }\href
  {\doibase 10.1103/PhysRevLett.93.072301} {\bibfield  {journal} {\bibinfo
  {journal} {Phys. Rev. Lett.}\ }\textbf {\bibinfo {volume} {93}},\ \bibinfo
  {pages} {072301} (\bibinfo {year} {2004})},\ \Eprint
  {http://arxiv.org/abs/nucl-th/0309040} {arXiv:nucl-th/0309040} \BibitemShut
  {NoStop}%
\bibitem [{\citenamefont {Zhang}\ \emph {et~al.}(2005)\citenamefont {Zhang},
  \citenamefont {Wang},\ and\ \citenamefont {Wang}}]{Zhang:2004qm}%
  \BibitemOpen
  \bibfield  {author} {\bibinfo {author} {\bibfnamefont {B.-W.}\ \bibnamefont
  {Zhang}}, \bibinfo {author} {\bibfnamefont {E.-k.}\ \bibnamefont {Wang}}, \
  and\ \bibinfo {author} {\bibfnamefont {X.-N.}\ \bibnamefont {Wang}},\ }\href
  {\doibase 10.1016/j.nuclphysa.2005.04.022} {\bibfield  {journal} {\bibinfo
  {journal} {Nucl. Phys. A}\ }\textbf {\bibinfo {volume} {757}},\ \bibinfo
  {pages} {493} (\bibinfo {year} {2005})},\ \Eprint
  {http://arxiv.org/abs/hep-ph/0412060} {arXiv:hep-ph/0412060} \BibitemShut
  {NoStop}%
\bibitem [{\citenamefont {Li}\ and\ \citenamefont
  {Vitev}(2019{\natexlab{a}})}]{Li:2018xuv}%
  \BibitemOpen
  \bibfield  {author} {\bibinfo {author} {\bibfnamefont {H.~T.}\ \bibnamefont
  {Li}}\ and\ \bibinfo {author} {\bibfnamefont {I.}~\bibnamefont {Vitev}},\
  }\href {\doibase 10.1007/JHEP07(2019)148} {\bibfield  {journal} {\bibinfo
  {journal} {JHEP}\ }\textbf {\bibinfo {volume} {07}},\ \bibinfo {pages} {148}
  (\bibinfo {year} {2019}{\natexlab{a}})},\ \Eprint
  {http://arxiv.org/abs/1811.07905} {arXiv:1811.07905 [hep-ph]} \BibitemShut
  {NoStop}%
\bibitem [{\citenamefont {Kang}\ \emph {et~al.}(2017)\citenamefont {Kang},
  \citenamefont {Ringer},\ and\ \citenamefont {Vitev}}]{Kang:2016ofv}%
  \BibitemOpen
  \bibfield  {author} {\bibinfo {author} {\bibfnamefont {Z.-B.}\ \bibnamefont
  {Kang}}, \bibinfo {author} {\bibfnamefont {F.}~\bibnamefont {Ringer}}, \ and\
  \bibinfo {author} {\bibfnamefont {I.}~\bibnamefont {Vitev}},\ }\href
  {\doibase 10.1007/JHEP03(2017)146} {\bibfield  {journal} {\bibinfo  {journal}
  {JHEP}\ }\textbf {\bibinfo {volume} {03}},\ \bibinfo {pages} {146} (\bibinfo
  {year} {2017})},\ \Eprint {http://arxiv.org/abs/1610.02043} {arXiv:1610.02043
  [hep-ph]} \BibitemShut {NoStop}%
\bibitem [{\citenamefont {Xing}\ \emph {et~al.}(2023)\citenamefont {Xing},
  \citenamefont {Qin},\ and\ \citenamefont {Cao}}]{Xing:2021xwc}%
  \BibitemOpen
  \bibfield  {author} {\bibinfo {author} {\bibfnamefont {W.-J.}\ \bibnamefont
  {Xing}}, \bibinfo {author} {\bibfnamefont {G.-Y.}\ \bibnamefont {Qin}}, \
  and\ \bibinfo {author} {\bibfnamefont {S.}~\bibnamefont {Cao}},\ }\href
  {\doibase 10.1016/j.physletb.2023.137733} {\bibfield  {journal} {\bibinfo
  {journal} {Phys. Lett. B}\ }\textbf {\bibinfo {volume} {838}},\ \bibinfo
  {pages} {137733} (\bibinfo {year} {2023})},\ \Eprint
  {http://arxiv.org/abs/2112.15062} {arXiv:2112.15062 [hep-ph]} \BibitemShut
  {NoStop}%
\bibitem [{\citenamefont {Huang}\ \emph {et~al.}(2013)\citenamefont {Huang},
  \citenamefont {Kang},\ and\ \citenamefont {Vitev}}]{Huang:2013vaa}%
  \BibitemOpen
  \bibfield  {author} {\bibinfo {author} {\bibfnamefont {J.}~\bibnamefont
  {Huang}}, \bibinfo {author} {\bibfnamefont {Z.-B.}\ \bibnamefont {Kang}}, \
  and\ \bibinfo {author} {\bibfnamefont {I.}~\bibnamefont {Vitev}},\ }\href
  {\doibase 10.1016/j.physletb.2013.08.009} {\bibfield  {journal} {\bibinfo
  {journal} {Phys. Lett. B}\ }\textbf {\bibinfo {volume} {726}},\ \bibinfo
  {pages} {251} (\bibinfo {year} {2013})},\ \Eprint
  {http://arxiv.org/abs/1306.0909} {arXiv:1306.0909 [hep-ph]} \BibitemShut
  {NoStop}%
\bibitem [{\citenamefont {Kang}\ \emph {et~al.}(2019)\citenamefont {Kang},
  \citenamefont {Reiten}, \citenamefont {Vitev},\ and\ \citenamefont
  {Yoon}}]{Kang:2018wrs}%
  \BibitemOpen
  \bibfield  {author} {\bibinfo {author} {\bibfnamefont {Z.-B.}\ \bibnamefont
  {Kang}}, \bibinfo {author} {\bibfnamefont {J.}~\bibnamefont {Reiten}},
  \bibinfo {author} {\bibfnamefont {I.}~\bibnamefont {Vitev}}, \ and\ \bibinfo
  {author} {\bibfnamefont {B.}~\bibnamefont {Yoon}},\ }\href {\doibase
  10.1103/PhysRevD.99.034006} {\bibfield  {journal} {\bibinfo  {journal} {Phys.
  Rev. D}\ }\textbf {\bibinfo {volume} {99}},\ \bibinfo {pages} {034006}
  (\bibinfo {year} {2019})},\ \Eprint {http://arxiv.org/abs/1810.10007}
  {arXiv:1810.10007 [hep-ph]} \BibitemShut {NoStop}%
\bibitem [{\citenamefont {Li}\ and\ \citenamefont
  {Vitev}(2019{\natexlab{b}})}]{Li:2017wwc}%
  \BibitemOpen
  \bibfield  {author} {\bibinfo {author} {\bibfnamefont {H.~T.}\ \bibnamefont
  {Li}}\ and\ \bibinfo {author} {\bibfnamefont {I.}~\bibnamefont {Vitev}},\
  }\href {\doibase 10.1016/j.physletb.2019.04.052} {\bibfield  {journal}
  {\bibinfo  {journal} {Phys. Lett. B}\ }\textbf {\bibinfo {volume} {793}},\
  \bibinfo {pages} {259} (\bibinfo {year} {2019}{\natexlab{b}})},\ \Eprint
  {http://arxiv.org/abs/1801.00008} {arXiv:1801.00008 [hep-ph]} \BibitemShut
  {NoStop}%
\bibitem [{\citenamefont {Cao}\ \emph {et~al.}(2019)\citenamefont {Cao} \emph
  {et~al.}}]{Cao:2018ews}%
  \BibitemOpen
  \bibfield  {author} {\bibinfo {author} {\bibfnamefont {S.}~\bibnamefont
  {Cao}} \emph {et~al.},\ }\href {\doibase 10.1103/PhysRevC.99.054907}
  {\bibfield  {journal} {\bibinfo  {journal} {Phys. Rev. C}\ }\textbf {\bibinfo
  {volume} {99}},\ \bibinfo {pages} {054907} (\bibinfo {year} {2019})},\
  \Eprint {http://arxiv.org/abs/1809.07894} {arXiv:1809.07894 [nucl-th]}
  \BibitemShut {NoStop}%
\bibitem [{\citenamefont {Andronic}\ \emph {et~al.}(2016)\citenamefont
  {Andronic} \emph {et~al.}}]{Andronic:2015wma}%
  \BibitemOpen
  \bibfield  {author} {\bibinfo {author} {\bibfnamefont {A.}~\bibnamefont
  {Andronic}} \emph {et~al.},\ }\href {\doibase 10.1140/epjc/s10052-015-3819-5}
  {\bibfield  {journal} {\bibinfo  {journal} {Eur. Phys. J. C}\ }\textbf
  {\bibinfo {volume} {76}},\ \bibinfo {pages} {107} (\bibinfo {year} {2016})},\
  \Eprint {http://arxiv.org/abs/1506.03981} {arXiv:1506.03981 [nucl-ex]}
  \BibitemShut {NoStop}%
\bibitem [{\citenamefont {Beraudo}\ \emph {et~al.}(2018)\citenamefont {Beraudo}
  \emph {et~al.}}]{Rapp:2018qla}%
  \BibitemOpen
  \bibfield  {author} {\bibinfo {author} {\bibfnamefont {A.}~\bibnamefont
  {Beraudo}} \emph {et~al.},\ }\href {\doibase 10.1016/j.nuclphysa.2018.09.002}
  {\bibfield  {journal} {\bibinfo  {journal} {Nucl. Phys. A}\ }\textbf
  {\bibinfo {volume} {979}},\ \bibinfo {pages} {21} (\bibinfo {year} {2018})},\
  \Eprint {http://arxiv.org/abs/1803.03824} {arXiv:1803.03824 [nucl-th]}
  \BibitemShut {NoStop}%
\bibitem [{\citenamefont {Dong}\ \emph {et~al.}(2019)\citenamefont {Dong},
  \citenamefont {Lee},\ and\ \citenamefont {Rapp}}]{Dong:2019byy}%
  \BibitemOpen
  \bibfield  {author} {\bibinfo {author} {\bibfnamefont {X.}~\bibnamefont
  {Dong}}, \bibinfo {author} {\bibfnamefont {Y.-J.}\ \bibnamefont {Lee}}, \
  and\ \bibinfo {author} {\bibfnamefont {R.}~\bibnamefont {Rapp}},\ }\href
  {\doibase 10.1146/annurev-nucl-101918-023806} {\bibfield  {journal} {\bibinfo
   {journal} {Ann. Rev. Nucl. Part. Sci.}\ }\textbf {\bibinfo {volume} {69}},\
  \bibinfo {pages} {417} (\bibinfo {year} {2019})},\ \Eprint
  {http://arxiv.org/abs/1903.07709} {arXiv:1903.07709 [nucl-ex]} \BibitemShut
  {NoStop}%
\bibitem [{\citenamefont {Apolin\'ario}\ \emph {et~al.}(2022)\citenamefont
  {Apolin\'ario}, \citenamefont {Lee},\ and\ \citenamefont
  {Winn}}]{Apolinario:2022vzg}%
  \BibitemOpen
  \bibfield  {author} {\bibinfo {author} {\bibfnamefont {L.}~\bibnamefont
  {Apolin\'ario}}, \bibinfo {author} {\bibfnamefont {Y.-J.}\ \bibnamefont
  {Lee}}, \ and\ \bibinfo {author} {\bibfnamefont {M.}~\bibnamefont {Winn}},\
  }\href@noop {} {\  (\bibinfo {year} {2022})},\ \Eprint
  {http://arxiv.org/abs/2203.16352} {arXiv:2203.16352 [hep-ph]} \BibitemShut
  {NoStop}%
\bibitem [{\citenamefont {Sirunyan}\ \emph {et~al.}(2019)\citenamefont
  {Sirunyan} \emph {et~al.}}]{CMS:2018bwt}%
  \BibitemOpen
  \bibfield  {author} {\bibinfo {author} {\bibfnamefont {A.~M.}\ \bibnamefont
  {Sirunyan}} \emph {et~al.} (\bibinfo {collaboration} {CMS}),\ }\href
  {\doibase 10.1103/PhysRevLett.123.022001} {\bibfield  {journal} {\bibinfo
  {journal} {Phys. Rev. Lett.}\ }\textbf {\bibinfo {volume} {123}},\ \bibinfo
  {pages} {022001} (\bibinfo {year} {2019})},\ \Eprint
  {http://arxiv.org/abs/1810.11102} {arXiv:1810.11102 [hep-ex]} \BibitemShut
  {NoStop}%
\bibitem [{\citenamefont {Chatrchyan}\ \emph {et~al.}(2014)\citenamefont
  {Chatrchyan} \emph {et~al.}}]{CMS:2013qak}%
  \BibitemOpen
  \bibfield  {author} {\bibinfo {author} {\bibfnamefont {S.}~\bibnamefont
  {Chatrchyan}} \emph {et~al.} (\bibinfo {collaboration} {CMS}),\ }\href
  {\doibase 10.1103/PhysRevLett.113.132301} {\bibfield  {journal} {\bibinfo
  {journal} {Phys. Rev. Lett.}\ }\textbf {\bibinfo {volume} {113}},\ \bibinfo
  {pages} {132301} (\bibinfo {year} {2014})},\ \bibinfo {note} {[Erratum:
  Phys.Rev.Lett. 115, 029903 (2015)]},\ \Eprint
  {http://arxiv.org/abs/1312.4198} {arXiv:1312.4198 [nucl-ex]} \BibitemShut
  {NoStop}%
\bibitem [{\citenamefont {Sirunyan}\ \emph {et~al.}(2018)\citenamefont
  {Sirunyan} \emph {et~al.}}]{CMS:2018dqf}%
  \BibitemOpen
  \bibfield  {author} {\bibinfo {author} {\bibfnamefont {A.~M.}\ \bibnamefont
  {Sirunyan}} \emph {et~al.} (\bibinfo {collaboration} {CMS}),\ }\href
  {\doibase 10.1007/JHEP03(2018)181} {\bibfield  {journal} {\bibinfo  {journal}
  {JHEP}\ }\textbf {\bibinfo {volume} {03}},\ \bibinfo {pages} {181} (\bibinfo
  {year} {2018})},\ \Eprint {http://arxiv.org/abs/1802.00707} {arXiv:1802.00707
  [hep-ex]} \BibitemShut {NoStop}%
\bibitem [{\citenamefont {Khachatryan}\ \emph {et~al.}(2016)\citenamefont
  {Khachatryan} \emph {et~al.}}]{CMS:2015sfx}%
  \BibitemOpen
  \bibfield  {author} {\bibinfo {author} {\bibfnamefont {V.}~\bibnamefont
  {Khachatryan}} \emph {et~al.} (\bibinfo {collaboration} {CMS}),\ }\href
  {\doibase 10.1103/PhysRevLett.116.032301} {\bibfield  {journal} {\bibinfo
  {journal} {Phys. Rev. Lett.}\ }\textbf {\bibinfo {volume} {116}},\ \bibinfo
  {pages} {032301} (\bibinfo {year} {2016})},\ \Eprint
  {http://arxiv.org/abs/1508.06678} {arXiv:1508.06678 [nucl-ex]} \BibitemShut
  {NoStop}%
\bibitem [{\citenamefont {CMS}(2022)}]{CMS:2022btc}%
  \BibitemOpen
  \bibfield  {author} {\bibinfo {author} {\bibnamefont {CMS}},\ }\href@noop {}
  {\  (\bibinfo {year} {2022})},\ \Eprint {http://arxiv.org/abs/2210.08547}
  {arXiv:2210.08547 [hep-ex]} \BibitemShut {NoStop}%
\bibitem [{\citenamefont {Dokshitzer}\ \emph {et~al.}(1991)\citenamefont
  {Dokshitzer}, \citenamefont {Khoze},\ and\ \citenamefont
  {Troian}}]{Dokshitzer:1991fd}%
  \BibitemOpen
  \bibfield  {author} {\bibinfo {author} {\bibfnamefont {Y.~L.}\ \bibnamefont
  {Dokshitzer}}, \bibinfo {author} {\bibfnamefont {V.~A.}\ \bibnamefont
  {Khoze}}, \ and\ \bibinfo {author} {\bibfnamefont {S.~I.}\ \bibnamefont
  {Troian}},\ }\href {\doibase 10.1088/0954-3899/17/10/023} {\bibfield
  {journal} {\bibinfo  {journal} {J. Phys. G}\ }\textbf {\bibinfo {volume}
  {17}},\ \bibinfo {pages} {1602} (\bibinfo {year} {1991})}\BibitemShut
  {NoStop}%
\bibitem [{\citenamefont {Acharya}\ \emph {et~al.}(2022)\citenamefont {Acharya}
  \emph {et~al.}}]{ALICE:2021aqk}%
  \BibitemOpen
  \bibfield  {author} {\bibinfo {author} {\bibfnamefont {S.}~\bibnamefont
  {Acharya}} \emph {et~al.} (\bibinfo {collaboration} {ALICE}),\ }\href
  {\doibase 10.1038/s41586-022-04572-w} {\bibfield  {journal} {\bibinfo
  {journal} {Nature}\ }\textbf {\bibinfo {volume} {605}},\ \bibinfo {pages}
  {440} (\bibinfo {year} {2022})},\ \bibinfo {note} {[Erratum: Nature 607, E22
  (2022)]},\ \Eprint {http://arxiv.org/abs/2106.05713} {arXiv:2106.05713
  [nucl-ex]} \BibitemShut {NoStop}%
\bibitem [{\citenamefont {Cunqueiro}\ and\ \citenamefont
  {P\l{}osko\'n}(2019)}]{Cunqueiro:2018jbh}%
  \BibitemOpen
  \bibfield  {author} {\bibinfo {author} {\bibfnamefont {L.}~\bibnamefont
  {Cunqueiro}}\ and\ \bibinfo {author} {\bibfnamefont {M.}~\bibnamefont
  {P\l{}osko\'n}},\ }\href {\doibase 10.1103/PhysRevD.99.074027} {\bibfield
  {journal} {\bibinfo  {journal} {Phys. Rev. D}\ }\textbf {\bibinfo {volume}
  {99}},\ \bibinfo {pages} {074027} (\bibinfo {year} {2019})},\ \Eprint
  {http://arxiv.org/abs/1812.00102} {arXiv:1812.00102 [hep-ph]} \BibitemShut
  {NoStop}%
\bibitem [{\citenamefont {Mehtar-Tani}\ \emph {et~al.}(2011)\citenamefont
  {Mehtar-Tani}, \citenamefont {Salgado},\ and\ \citenamefont
  {Tywoniuk}}]{Mehtar-Tani:2010ebp}%
  \BibitemOpen
  \bibfield  {author} {\bibinfo {author} {\bibfnamefont {Y.}~\bibnamefont
  {Mehtar-Tani}}, \bibinfo {author} {\bibfnamefont {C.~A.}\ \bibnamefont
  {Salgado}}, \ and\ \bibinfo {author} {\bibfnamefont {K.}~\bibnamefont
  {Tywoniuk}},\ }\href {\doibase 10.1103/PhysRevLett.106.122002} {\bibfield
  {journal} {\bibinfo  {journal} {Phys. Rev. Lett.}\ }\textbf {\bibinfo
  {volume} {106}},\ \bibinfo {pages} {122002} (\bibinfo {year} {2011})},\
  \Eprint {http://arxiv.org/abs/1009.2965} {arXiv:1009.2965 [hep-ph]}
  \BibitemShut {NoStop}%
\bibitem [{\citenamefont {Mulligan}\ and\ \citenamefont
  {Ploskon}(2020)}]{Mulligan:2020tim}%
  \BibitemOpen
  \bibfield  {author} {\bibinfo {author} {\bibfnamefont {J.}~\bibnamefont
  {Mulligan}}\ and\ \bibinfo {author} {\bibfnamefont {M.}~\bibnamefont
  {Ploskon}},\ }\href {\doibase 10.1103/PhysRevC.102.044913} {\bibfield
  {journal} {\bibinfo  {journal} {Phys. Rev. C}\ }\textbf {\bibinfo {volume}
  {102}},\ \bibinfo {pages} {044913} (\bibinfo {year} {2020})},\ \Eprint
  {http://arxiv.org/abs/2006.01812} {arXiv:2006.01812 [hep-ph]} \BibitemShut
  {NoStop}%
\bibitem [{\citenamefont {Cunqueiro}\ \emph {et~al.}(2023)\citenamefont
  {Cunqueiro}, \citenamefont {Napoletano},\ and\ \citenamefont
  {Soto-Ontoso}}]{Cunqueiro:2022svx}%
  \BibitemOpen
  \bibfield  {author} {\bibinfo {author} {\bibfnamefont {L.}~\bibnamefont
  {Cunqueiro}}, \bibinfo {author} {\bibfnamefont {D.}~\bibnamefont
  {Napoletano}}, \ and\ \bibinfo {author} {\bibfnamefont {A.}~\bibnamefont
  {Soto-Ontoso}},\ }\href {\doibase 10.1103/PhysRevD.107.094008} {\bibfield
  {journal} {\bibinfo  {journal} {Phys. Rev. D}\ }\textbf {\bibinfo {volume}
  {107}},\ \bibinfo {pages} {094008} (\bibinfo {year} {2023})},\ \Eprint
  {http://arxiv.org/abs/2211.11789} {arXiv:2211.11789 [hep-ph]} \BibitemShut
  {NoStop}%
\bibitem [{\citenamefont {Basham}\ \emph
  {et~al.}(1979{\natexlab{a}})\citenamefont {Basham}, \citenamefont {Brown},
  \citenamefont {Ellis},\ and\ \citenamefont {Love}}]{Basham:1979gh}%
  \BibitemOpen
  \bibfield  {author} {\bibinfo {author} {\bibfnamefont {C.~L.}\ \bibnamefont
  {Basham}}, \bibinfo {author} {\bibfnamefont {L.~S.}\ \bibnamefont {Brown}},
  \bibinfo {author} {\bibfnamefont {S.~D.}\ \bibnamefont {Ellis}}, \ and\
  \bibinfo {author} {\bibfnamefont {S.~T.}\ \bibnamefont {Love}},\ }\href
  {\doibase 10.1016/0370-2693(79)90601-4} {\bibfield  {journal} {\bibinfo
  {journal} {Phys. Lett. B}\ }\textbf {\bibinfo {volume} {85}},\ \bibinfo
  {pages} {297} (\bibinfo {year} {1979}{\natexlab{a}})}\BibitemShut {NoStop}%
\bibitem [{\citenamefont {Basham}\ \emph
  {et~al.}(1979{\natexlab{b}})\citenamefont {Basham}, \citenamefont {Brown},
  \citenamefont {Ellis},\ and\ \citenamefont {Love}}]{Basham:1978zq}%
  \BibitemOpen
  \bibfield  {author} {\bibinfo {author} {\bibfnamefont {C.}~\bibnamefont
  {Basham}}, \bibinfo {author} {\bibfnamefont {L.}~\bibnamefont {Brown}},
  \bibinfo {author} {\bibfnamefont {S.}~\bibnamefont {Ellis}}, \ and\ \bibinfo
  {author} {\bibfnamefont {S.}~\bibnamefont {Love}},\ }\href {\doibase
  10.1103/PhysRevD.19.2018} {\bibfield  {journal} {\bibinfo  {journal} {Phys.
  Rev. D}\ }\textbf {\bibinfo {volume} {19}},\ \bibinfo {pages} {2018}
  (\bibinfo {year} {1979}{\natexlab{b}})}\BibitemShut {NoStop}%
\bibitem [{\citenamefont {Basham}\ \emph
  {et~al.}(1978{\natexlab{a}})\citenamefont {Basham}, \citenamefont {Brown},
  \citenamefont {Ellis},\ and\ \citenamefont {Love}}]{Basham:1978bw}%
  \BibitemOpen
  \bibfield  {author} {\bibinfo {author} {\bibfnamefont {C.}~\bibnamefont
  {Basham}}, \bibinfo {author} {\bibfnamefont {L.~S.}\ \bibnamefont {Brown}},
  \bibinfo {author} {\bibfnamefont {S.~D.}\ \bibnamefont {Ellis}}, \ and\
  \bibinfo {author} {\bibfnamefont {S.~T.}\ \bibnamefont {Love}},\ }\href
  {\doibase 10.1103/PhysRevLett.41.1585} {\bibfield  {journal} {\bibinfo
  {journal} {Phys. Rev. Lett.}\ }\textbf {\bibinfo {volume} {41}},\ \bibinfo
  {pages} {1585} (\bibinfo {year} {1978}{\natexlab{a}})}\BibitemShut {NoStop}%
\bibitem [{\citenamefont {Basham}\ \emph
  {et~al.}(1978{\natexlab{b}})\citenamefont {Basham}, \citenamefont {Brown},
  \citenamefont {Ellis},\ and\ \citenamefont {Love}}]{Basham:1977iq}%
  \BibitemOpen
  \bibfield  {author} {\bibinfo {author} {\bibfnamefont {C.~L.}\ \bibnamefont
  {Basham}}, \bibinfo {author} {\bibfnamefont {L.~S.}\ \bibnamefont {Brown}},
  \bibinfo {author} {\bibfnamefont {S.~D.}\ \bibnamefont {Ellis}}, \ and\
  \bibinfo {author} {\bibfnamefont {S.~T.}\ \bibnamefont {Love}},\ }\href
  {\doibase 10.1103/PhysRevD.17.2298} {\bibfield  {journal} {\bibinfo
  {journal} {Phys. Rev. D}\ }\textbf {\bibinfo {volume} {17}},\ \bibinfo
  {pages} {2298} (\bibinfo {year} {1978}{\natexlab{b}})}\BibitemShut {NoStop}%
\bibitem [{\citenamefont {Hofman}\ and\ \citenamefont
  {Maldacena}(2008)}]{Hofman:2008ar}%
  \BibitemOpen
  \bibfield  {author} {\bibinfo {author} {\bibfnamefont {D.~M.}\ \bibnamefont
  {Hofman}}\ and\ \bibinfo {author} {\bibfnamefont {J.}~\bibnamefont
  {Maldacena}},\ }\href {\doibase 10.1088/1126-6708/2008/05/012} {\bibfield
  {journal} {\bibinfo  {journal} {JHEP}\ }\textbf {\bibinfo {volume} {05}},\
  \bibinfo {pages} {012} (\bibinfo {year} {2008})},\ \Eprint
  {http://arxiv.org/abs/0803.1467} {arXiv:0803.1467 [hep-th]} \BibitemShut
  {NoStop}%
\bibitem [{\citenamefont {Sveshnikov}\ and\ \citenamefont
  {Tkachov}(1996)}]{Sveshnikov:1995vi}%
  \BibitemOpen
  \bibfield  {author} {\bibinfo {author} {\bibfnamefont {N.}~\bibnamefont
  {Sveshnikov}}\ and\ \bibinfo {author} {\bibfnamefont {F.}~\bibnamefont
  {Tkachov}},\ }\href {\doibase 10.1016/0370-2693(96)00558-8} {\bibfield
  {journal} {\bibinfo  {journal} {Phys. Lett. B}\ }\textbf {\bibinfo {volume}
  {382}},\ \bibinfo {pages} {403} (\bibinfo {year} {1996})},\ \Eprint
  {http://arxiv.org/abs/hep-ph/9512370} {arXiv:hep-ph/9512370} \BibitemShut
  {NoStop}%
\bibitem [{\citenamefont {Tkachov}(1997)}]{Tkachov:1995kk}%
  \BibitemOpen
  \bibfield  {author} {\bibinfo {author} {\bibfnamefont {F.~V.}\ \bibnamefont
  {Tkachov}},\ }\href {\doibase 10.1142/S0217751X97002899} {\bibfield
  {journal} {\bibinfo  {journal} {Int. J. Mod. Phys. A}\ }\textbf {\bibinfo
  {volume} {12}},\ \bibinfo {pages} {5411} (\bibinfo {year} {1997})},\ \Eprint
  {http://arxiv.org/abs/hep-ph/9601308} {arXiv:hep-ph/9601308} \BibitemShut
  {NoStop}%
\bibitem [{\citenamefont {Korchemsky}\ and\ \citenamefont
  {Sterman}(1999)}]{Korchemsky:1999kt}%
  \BibitemOpen
  \bibfield  {author} {\bibinfo {author} {\bibfnamefont {G.~P.}\ \bibnamefont
  {Korchemsky}}\ and\ \bibinfo {author} {\bibfnamefont {G.~F.}\ \bibnamefont
  {Sterman}},\ }\href {\doibase 10.1016/S0550-3213(99)00308-9} {\bibfield
  {journal} {\bibinfo  {journal} {Nucl. Phys. B}\ }\textbf {\bibinfo {volume}
  {555}},\ \bibinfo {pages} {335} (\bibinfo {year} {1999})},\ \Eprint
  {http://arxiv.org/abs/hep-ph/9902341} {arXiv:hep-ph/9902341} \BibitemShut
  {NoStop}%
\bibitem [{\citenamefont {Bauer}\ \emph {et~al.}(2008)\citenamefont {Bauer},
  \citenamefont {Fleming}, \citenamefont {Lee},\ and\ \citenamefont
  {Sterman}}]{Bauer:2008dt}%
  \BibitemOpen
  \bibfield  {author} {\bibinfo {author} {\bibfnamefont {C.~W.}\ \bibnamefont
  {Bauer}}, \bibinfo {author} {\bibfnamefont {S.~P.}\ \bibnamefont {Fleming}},
  \bibinfo {author} {\bibfnamefont {C.}~\bibnamefont {Lee}}, \ and\ \bibinfo
  {author} {\bibfnamefont {G.~F.}\ \bibnamefont {Sterman}},\ }\href {\doibase
  10.1103/PhysRevD.78.034027} {\bibfield  {journal} {\bibinfo  {journal} {Phys.
  Rev. D}\ }\textbf {\bibinfo {volume} {78}},\ \bibinfo {pages} {034027}
  (\bibinfo {year} {2008})},\ \Eprint {http://arxiv.org/abs/0801.4569}
  {arXiv:0801.4569 [hep-ph]} \BibitemShut {NoStop}%
\bibitem [{\citenamefont {Belitsky}\ \emph
  {et~al.}(2014{\natexlab{a}})\citenamefont {Belitsky}, \citenamefont
  {Hohenegger}, \citenamefont {Korchemsky}, \citenamefont {Sokatchev},\ and\
  \citenamefont {Zhiboedov}}]{Belitsky:2013xxa}%
  \BibitemOpen
  \bibfield  {author} {\bibinfo {author} {\bibfnamefont {A.}~\bibnamefont
  {Belitsky}}, \bibinfo {author} {\bibfnamefont {S.}~\bibnamefont
  {Hohenegger}}, \bibinfo {author} {\bibfnamefont {G.}~\bibnamefont
  {Korchemsky}}, \bibinfo {author} {\bibfnamefont {E.}~\bibnamefont
  {Sokatchev}}, \ and\ \bibinfo {author} {\bibfnamefont {A.}~\bibnamefont
  {Zhiboedov}},\ }\href {\doibase 10.1016/j.nuclphysb.2014.04.020} {\bibfield
  {journal} {\bibinfo  {journal} {Nucl. Phys. B}\ }\textbf {\bibinfo {volume}
  {884}},\ \bibinfo {pages} {305} (\bibinfo {year} {2014}{\natexlab{a}})},\
  \Eprint {http://arxiv.org/abs/1309.0769} {arXiv:1309.0769 [hep-th]}
  \BibitemShut {NoStop}%
\bibitem [{\citenamefont {Belitsky}\ \emph
  {et~al.}(2014{\natexlab{b}})\citenamefont {Belitsky}, \citenamefont
  {Hohenegger}, \citenamefont {Korchemsky}, \citenamefont {Sokatchev},\ and\
  \citenamefont {Zhiboedov}}]{Belitsky:2013bja}%
  \BibitemOpen
  \bibfield  {author} {\bibinfo {author} {\bibfnamefont {A.}~\bibnamefont
  {Belitsky}}, \bibinfo {author} {\bibfnamefont {S.}~\bibnamefont
  {Hohenegger}}, \bibinfo {author} {\bibfnamefont {G.}~\bibnamefont
  {Korchemsky}}, \bibinfo {author} {\bibfnamefont {E.}~\bibnamefont
  {Sokatchev}}, \ and\ \bibinfo {author} {\bibfnamefont {A.}~\bibnamefont
  {Zhiboedov}},\ }\href {\doibase 10.1016/j.nuclphysb.2014.04.019} {\bibfield
  {journal} {\bibinfo  {journal} {Nucl. Phys. B}\ }\textbf {\bibinfo {volume}
  {884}},\ \bibinfo {pages} {206} (\bibinfo {year} {2014}{\natexlab{b}})},\
  \Eprint {http://arxiv.org/abs/1309.1424} {arXiv:1309.1424 [hep-th]}
  \BibitemShut {NoStop}%
\bibitem [{\citenamefont {Kravchuk}\ and\ \citenamefont
  {Simmons-Duffin}(2018)}]{Kravchuk:2018htv}%
  \BibitemOpen
  \bibfield  {author} {\bibinfo {author} {\bibfnamefont {P.}~\bibnamefont
  {Kravchuk}}\ and\ \bibinfo {author} {\bibfnamefont {D.}~\bibnamefont
  {Simmons-Duffin}},\ }\href {\doibase 10.1007/JHEP11(2018)102} {\bibfield
  {journal} {\bibinfo  {journal} {JHEP}\ }\textbf {\bibinfo {volume} {11}},\
  \bibinfo {pages} {102} (\bibinfo {year} {2018})},\ \Eprint
  {http://arxiv.org/abs/1805.00098} {arXiv:1805.00098 [hep-th]} \BibitemShut
  {NoStop}%
\bibitem [{\citenamefont {Belitsky}\ \emph
  {et~al.}(2014{\natexlab{c}})\citenamefont {Belitsky}, \citenamefont
  {Hohenegger}, \citenamefont {Korchemsky}, \citenamefont {Sokatchev},\ and\
  \citenamefont {Zhiboedov}}]{Belitsky:2013ofa}%
  \BibitemOpen
  \bibfield  {author} {\bibinfo {author} {\bibfnamefont {A.}~\bibnamefont
  {Belitsky}}, \bibinfo {author} {\bibfnamefont {S.}~\bibnamefont
  {Hohenegger}}, \bibinfo {author} {\bibfnamefont {G.}~\bibnamefont
  {Korchemsky}}, \bibinfo {author} {\bibfnamefont {E.}~\bibnamefont
  {Sokatchev}}, \ and\ \bibinfo {author} {\bibfnamefont {A.}~\bibnamefont
  {Zhiboedov}},\ }\href {\doibase 10.1103/PhysRevLett.112.071601} {\bibfield
  {journal} {\bibinfo  {journal} {Phys. Rev. Lett.}\ }\textbf {\bibinfo
  {volume} {112}},\ \bibinfo {pages} {071601} (\bibinfo {year}
  {2014}{\natexlab{c}})},\ \Eprint {http://arxiv.org/abs/1311.6800}
  {arXiv:1311.6800 [hep-th]} \BibitemShut {NoStop}%
\bibitem [{\citenamefont {Korchemsky}\ and\ \citenamefont
  {Sokatchev}(2015)}]{Korchemsky:2015ssa}%
  \BibitemOpen
  \bibfield  {author} {\bibinfo {author} {\bibfnamefont {G.}~\bibnamefont
  {Korchemsky}}\ and\ \bibinfo {author} {\bibfnamefont {E.}~\bibnamefont
  {Sokatchev}},\ }\href {\doibase 10.1007/JHEP12(2015)133} {\bibfield
  {journal} {\bibinfo  {journal} {JHEP}\ }\textbf {\bibinfo {volume} {12}},\
  \bibinfo {pages} {133} (\bibinfo {year} {2015})},\ \Eprint
  {http://arxiv.org/abs/1504.07904} {arXiv:1504.07904 [hep-th]} \BibitemShut
  {NoStop}%
\bibitem [{\citenamefont {Belitsky}\ \emph {et~al.}(2016)\citenamefont
  {Belitsky}, \citenamefont {Hohenegger}, \citenamefont {Korchemsky},\ and\
  \citenamefont {Sokatchev}}]{Belitsky:2014zha}%
  \BibitemOpen
  \bibfield  {author} {\bibinfo {author} {\bibfnamefont {A.}~\bibnamefont
  {Belitsky}}, \bibinfo {author} {\bibfnamefont {S.}~\bibnamefont
  {Hohenegger}}, \bibinfo {author} {\bibfnamefont {G.}~\bibnamefont
  {Korchemsky}}, \ and\ \bibinfo {author} {\bibfnamefont {E.}~\bibnamefont
  {Sokatchev}},\ }\href {\doibase 10.1016/j.nuclphysb.2016.01.008} {\bibfield
  {journal} {\bibinfo  {journal} {Nucl. Phys. B}\ }\textbf {\bibinfo {volume}
  {904}},\ \bibinfo {pages} {176} (\bibinfo {year} {2016})},\ \Eprint
  {http://arxiv.org/abs/1409.2502} {arXiv:1409.2502 [hep-th]} \BibitemShut
  {NoStop}%
\bibitem [{\citenamefont {Dixon}\ \emph {et~al.}(2018)\citenamefont {Dixon},
  \citenamefont {Luo}, \citenamefont {Shtabovenko}, \citenamefont {Yang},\ and\
  \citenamefont {Zhu}}]{Dixon:2018qgp}%
  \BibitemOpen
  \bibfield  {author} {\bibinfo {author} {\bibfnamefont {L.~J.}\ \bibnamefont
  {Dixon}}, \bibinfo {author} {\bibfnamefont {M.-X.}\ \bibnamefont {Luo}},
  \bibinfo {author} {\bibfnamefont {V.}~\bibnamefont {Shtabovenko}}, \bibinfo
  {author} {\bibfnamefont {T.-Z.}\ \bibnamefont {Yang}}, \ and\ \bibinfo
  {author} {\bibfnamefont {H.~X.}\ \bibnamefont {Zhu}},\ }\href {\doibase
  10.1103/PhysRevLett.120.102001} {\bibfield  {journal} {\bibinfo  {journal}
  {Phys. Rev. Lett.}\ }\textbf {\bibinfo {volume} {120}},\ \bibinfo {pages}
  {102001} (\bibinfo {year} {2018})},\ \Eprint
  {http://arxiv.org/abs/1801.03219} {arXiv:1801.03219 [hep-ph]} \BibitemShut
  {NoStop}%
\bibitem [{\citenamefont {Luo}\ \emph {et~al.}(2019)\citenamefont {Luo},
  \citenamefont {Shtabovenko}, \citenamefont {Yang},\ and\ \citenamefont
  {Zhu}}]{Luo:2019nig}%
  \BibitemOpen
  \bibfield  {author} {\bibinfo {author} {\bibfnamefont {M.-X.}\ \bibnamefont
  {Luo}}, \bibinfo {author} {\bibfnamefont {V.}~\bibnamefont {Shtabovenko}},
  \bibinfo {author} {\bibfnamefont {T.-Z.}\ \bibnamefont {Yang}}, \ and\
  \bibinfo {author} {\bibfnamefont {H.~X.}\ \bibnamefont {Zhu}},\ }\href
  {\doibase 10.1007/JHEP06(2019)037} {\bibfield  {journal} {\bibinfo  {journal}
  {JHEP}\ }\textbf {\bibinfo {volume} {06}},\ \bibinfo {pages} {037} (\bibinfo
  {year} {2019})},\ \Eprint {http://arxiv.org/abs/1903.07277} {arXiv:1903.07277
  [hep-ph]} \BibitemShut {NoStop}%
\bibitem [{\citenamefont {Henn}\ \emph {et~al.}(2019)\citenamefont {Henn},
  \citenamefont {Sokatchev}, \citenamefont {Yan},\ and\ \citenamefont
  {Zhiboedov}}]{Henn:2019gkr}%
  \BibitemOpen
  \bibfield  {author} {\bibinfo {author} {\bibfnamefont {J.}~\bibnamefont
  {Henn}}, \bibinfo {author} {\bibfnamefont {E.}~\bibnamefont {Sokatchev}},
  \bibinfo {author} {\bibfnamefont {K.}~\bibnamefont {Yan}}, \ and\ \bibinfo
  {author} {\bibfnamefont {A.}~\bibnamefont {Zhiboedov}},\ }\href {\doibase
  10.1103/PhysRevD.100.036010} {\bibfield  {journal} {\bibinfo  {journal}
  {Phys. Rev. D}\ }\textbf {\bibinfo {volume} {100}},\ \bibinfo {pages}
  {036010} (\bibinfo {year} {2019})},\ \Eprint
  {http://arxiv.org/abs/1903.05314} {arXiv:1903.05314 [hep-th]} \BibitemShut
  {NoStop}%
\bibitem [{\citenamefont {Chen}\ \emph
  {et~al.}(2020{\natexlab{a}})\citenamefont {Chen}, \citenamefont {Luo},
  \citenamefont {Moult}, \citenamefont {Yang}, \citenamefont {Zhang},\ and\
  \citenamefont {Zhu}}]{Chen:2019bpb}%
  \BibitemOpen
  \bibfield  {author} {\bibinfo {author} {\bibfnamefont {H.}~\bibnamefont
  {Chen}}, \bibinfo {author} {\bibfnamefont {M.-X.}\ \bibnamefont {Luo}},
  \bibinfo {author} {\bibfnamefont {I.}~\bibnamefont {Moult}}, \bibinfo
  {author} {\bibfnamefont {T.-Z.}\ \bibnamefont {Yang}}, \bibinfo {author}
  {\bibfnamefont {X.}~\bibnamefont {Zhang}}, \ and\ \bibinfo {author}
  {\bibfnamefont {H.~X.}\ \bibnamefont {Zhu}},\ }\href {\doibase
  10.1007/JHEP08(2020)028} {\bibfield  {journal} {\bibinfo  {journal} {JHEP}\
  }\textbf {\bibinfo {volume} {08}},\ \bibinfo {pages} {028} (\bibinfo {year}
  {2020}{\natexlab{a}})},\ \Eprint {http://arxiv.org/abs/1912.11050}
  {arXiv:1912.11050 [hep-ph]} \BibitemShut {NoStop}%
\bibitem [{\citenamefont {Dixon}\ \emph {et~al.}(2019)\citenamefont {Dixon},
  \citenamefont {Moult},\ and\ \citenamefont {Zhu}}]{Dixon:2019uzg}%
  \BibitemOpen
  \bibfield  {author} {\bibinfo {author} {\bibfnamefont {L.~J.}\ \bibnamefont
  {Dixon}}, \bibinfo {author} {\bibfnamefont {I.}~\bibnamefont {Moult}}, \ and\
  \bibinfo {author} {\bibfnamefont {H.~X.}\ \bibnamefont {Zhu}},\ }\href
  {\doibase 10.1103/PhysRevD.100.014009} {\bibfield  {journal} {\bibinfo
  {journal} {Phys. Rev. D}\ }\textbf {\bibinfo {volume} {100}},\ \bibinfo
  {pages} {014009} (\bibinfo {year} {2019})},\ \Eprint
  {http://arxiv.org/abs/1905.01310} {arXiv:1905.01310 [hep-ph]} \BibitemShut
  {NoStop}%
\bibitem [{\citenamefont {Korchemsky}(2020)}]{Korchemsky:2019nzm}%
  \BibitemOpen
  \bibfield  {author} {\bibinfo {author} {\bibfnamefont {G.}~\bibnamefont
  {Korchemsky}},\ }\href {\doibase 10.1007/JHEP01(2020)008} {\bibfield
  {journal} {\bibinfo  {journal} {JHEP}\ }\textbf {\bibinfo {volume} {01}},\
  \bibinfo {pages} {008} (\bibinfo {year} {2020})},\ \Eprint
  {http://arxiv.org/abs/1905.01444} {arXiv:1905.01444 [hep-th]} \BibitemShut
  {NoStop}%
\bibitem [{\citenamefont {Chicherin}\ \emph {et~al.}(2021)\citenamefont
  {Chicherin}, \citenamefont {Henn}, \citenamefont {Sokatchev},\ and\
  \citenamefont {Yan}}]{Chicherin:2020azt}%
  \BibitemOpen
  \bibfield  {author} {\bibinfo {author} {\bibfnamefont {D.}~\bibnamefont
  {Chicherin}}, \bibinfo {author} {\bibfnamefont {J.~M.}\ \bibnamefont {Henn}},
  \bibinfo {author} {\bibfnamefont {E.}~\bibnamefont {Sokatchev}}, \ and\
  \bibinfo {author} {\bibfnamefont {K.}~\bibnamefont {Yan}},\ }\href {\doibase
  10.1007/JHEP02(2021)053} {\bibfield  {journal} {\bibinfo  {journal} {JHEP}\
  }\textbf {\bibinfo {volume} {02}},\ \bibinfo {pages} {053} (\bibinfo {year}
  {2021})},\ \Eprint {http://arxiv.org/abs/2001.10806} {arXiv:2001.10806
  [hep-th]} \BibitemShut {NoStop}%
\bibitem [{\citenamefont {Kologlu}\ \emph {et~al.}(2020)\citenamefont
  {Kologlu}, \citenamefont {Kravchuk}, \citenamefont {Simmons-Duffin},\ and\
  \citenamefont {Zhiboedov}}]{Kologlu:2019bco}%
  \BibitemOpen
  \bibfield  {author} {\bibinfo {author} {\bibfnamefont {M.}~\bibnamefont
  {Kologlu}}, \bibinfo {author} {\bibfnamefont {P.}~\bibnamefont {Kravchuk}},
  \bibinfo {author} {\bibfnamefont {D.}~\bibnamefont {Simmons-Duffin}}, \ and\
  \bibinfo {author} {\bibfnamefont {A.}~\bibnamefont {Zhiboedov}},\ }\href
  {\doibase 10.1007/JHEP11(2020)096} {\bibfield  {journal} {\bibinfo  {journal}
  {JHEP}\ }\textbf {\bibinfo {volume} {11}},\ \bibinfo {pages} {096} (\bibinfo
  {year} {2020})},\ \Eprint {http://arxiv.org/abs/1904.05905} {arXiv:1904.05905
  [hep-th]} \BibitemShut {NoStop}%
\bibitem [{\citenamefont {Kologlu}\ \emph {et~al.}(2021)\citenamefont
  {Kologlu}, \citenamefont {Kravchuk}, \citenamefont {Simmons-Duffin},\ and\
  \citenamefont {Zhiboedov}}]{Kologlu:2019mfz}%
  \BibitemOpen
  \bibfield  {author} {\bibinfo {author} {\bibfnamefont {M.}~\bibnamefont
  {Kologlu}}, \bibinfo {author} {\bibfnamefont {P.}~\bibnamefont {Kravchuk}},
  \bibinfo {author} {\bibfnamefont {D.}~\bibnamefont {Simmons-Duffin}}, \ and\
  \bibinfo {author} {\bibfnamefont {A.}~\bibnamefont {Zhiboedov}},\ }\href
  {\doibase 10.1007/JHEP01(2021)128} {\bibfield  {journal} {\bibinfo  {journal}
  {JHEP}\ }\textbf {\bibinfo {volume} {01}},\ \bibinfo {pages} {128} (\bibinfo
  {year} {2021})},\ \Eprint {http://arxiv.org/abs/1905.01311} {arXiv:1905.01311
  [hep-th]} \BibitemShut {NoStop}%
\bibitem [{\citenamefont {Chang}\ \emph {et~al.}(2022)\citenamefont {Chang},
  \citenamefont {Kologlu}, \citenamefont {Kravchuk}, \citenamefont
  {Simmons-Duffin},\ and\ \citenamefont {Zhiboedov}}]{Chang:2020qpj}%
  \BibitemOpen
  \bibfield  {author} {\bibinfo {author} {\bibfnamefont {C.-H.}\ \bibnamefont
  {Chang}}, \bibinfo {author} {\bibfnamefont {M.}~\bibnamefont {Kologlu}},
  \bibinfo {author} {\bibfnamefont {P.}~\bibnamefont {Kravchuk}}, \bibinfo
  {author} {\bibfnamefont {D.}~\bibnamefont {Simmons-Duffin}}, \ and\ \bibinfo
  {author} {\bibfnamefont {A.}~\bibnamefont {Zhiboedov}},\ }\href {\doibase
  10.1007/JHEP05(2022)059} {\bibfield  {journal} {\bibinfo  {journal} {JHEP}\
  }\textbf {\bibinfo {volume} {05}},\ \bibinfo {pages} {059} (\bibinfo {year}
  {2022})},\ \Eprint {http://arxiv.org/abs/2010.04726} {arXiv:2010.04726
  [hep-th]} \BibitemShut {NoStop}%
\bibitem [{\citenamefont {Chen}\ \emph
  {et~al.}(2021{\natexlab{a}})\citenamefont {Chen}, \citenamefont {Yang},
  \citenamefont {Zhu},\ and\ \citenamefont {Zhu}}]{Chen:2020uvt}%
  \BibitemOpen
  \bibfield  {author} {\bibinfo {author} {\bibfnamefont {H.}~\bibnamefont
  {Chen}}, \bibinfo {author} {\bibfnamefont {T.-Z.}\ \bibnamefont {Yang}},
  \bibinfo {author} {\bibfnamefont {H.~X.}\ \bibnamefont {Zhu}}, \ and\
  \bibinfo {author} {\bibfnamefont {Y.~J.}\ \bibnamefont {Zhu}},\ }\href
  {\doibase 10.1088/1674-1137/abde2d} {\bibfield  {journal} {\bibinfo
  {journal} {Chin. Phys. C}\ }\textbf {\bibinfo {volume} {45}},\ \bibinfo
  {pages} {043101} (\bibinfo {year} {2021}{\natexlab{a}})},\ \Eprint
  {http://arxiv.org/abs/2006.10534} {arXiv:2006.10534 [hep-ph]} \BibitemShut
  {NoStop}%
\bibitem [{\citenamefont {Chen}\ \emph
  {et~al.}(2020{\natexlab{b}})\citenamefont {Chen}, \citenamefont {Moult},
  \citenamefont {Zhang},\ and\ \citenamefont {Zhu}}]{Chen:2020vvp}%
  \BibitemOpen
  \bibfield  {author} {\bibinfo {author} {\bibfnamefont {H.}~\bibnamefont
  {Chen}}, \bibinfo {author} {\bibfnamefont {I.}~\bibnamefont {Moult}},
  \bibinfo {author} {\bibfnamefont {X.}~\bibnamefont {Zhang}}, \ and\ \bibinfo
  {author} {\bibfnamefont {H.~X.}\ \bibnamefont {Zhu}},\ }\href {\doibase
  10.1103/PhysRevD.102.054012} {\bibfield  {journal} {\bibinfo  {journal}
  {Phys. Rev. D}\ }\textbf {\bibinfo {volume} {102}},\ \bibinfo {pages}
  {054012} (\bibinfo {year} {2020}{\natexlab{b}})},\ \Eprint
  {http://arxiv.org/abs/2004.11381} {arXiv:2004.11381 [hep-ph]} \BibitemShut
  {NoStop}%
\bibitem [{\citenamefont {Chen}\ \emph
  {et~al.}(2021{\natexlab{b}})\citenamefont {Chen}, \citenamefont {Moult},\
  and\ \citenamefont {Zhu}}]{Chen:2020adz}%
  \BibitemOpen
  \bibfield  {author} {\bibinfo {author} {\bibfnamefont {H.}~\bibnamefont
  {Chen}}, \bibinfo {author} {\bibfnamefont {I.}~\bibnamefont {Moult}}, \ and\
  \bibinfo {author} {\bibfnamefont {H.~X.}\ \bibnamefont {Zhu}},\ }\href
  {\doibase 10.1103/PhysRevLett.126.112003} {\bibfield  {journal} {\bibinfo
  {journal} {Phys. Rev. Lett.}\ }\textbf {\bibinfo {volume} {126}},\ \bibinfo
  {pages} {112003} (\bibinfo {year} {2021}{\natexlab{b}})},\ \Eprint
  {http://arxiv.org/abs/2011.02492} {arXiv:2011.02492 [hep-ph]} \BibitemShut
  {NoStop}%
\bibitem [{\citenamefont {Chen}\ \emph
  {et~al.}(2022{\natexlab{a}})\citenamefont {Chen}, \citenamefont {Moult},\
  and\ \citenamefont {Zhu}}]{Chen:2021gdk}%
  \BibitemOpen
  \bibfield  {author} {\bibinfo {author} {\bibfnamefont {H.}~\bibnamefont
  {Chen}}, \bibinfo {author} {\bibfnamefont {I.}~\bibnamefont {Moult}}, \ and\
  \bibinfo {author} {\bibfnamefont {H.~X.}\ \bibnamefont {Zhu}},\ }\href
  {\doibase 10.1007/JHEP08(2022)233} {\bibfield  {journal} {\bibinfo  {journal}
  {JHEP}\ }\textbf {\bibinfo {volume} {08}},\ \bibinfo {pages} {233} (\bibinfo
  {year} {2022}{\natexlab{a}})},\ \Eprint {http://arxiv.org/abs/2104.00009}
  {arXiv:2104.00009 [hep-ph]} \BibitemShut {NoStop}%
\bibitem [{\citenamefont {Korchemsky}\ \emph {et~al.}(2022)\citenamefont
  {Korchemsky}, \citenamefont {Sokatchev},\ and\ \citenamefont
  {Zhiboedov}}]{Korchemsky:2021okt}%
  \BibitemOpen
  \bibfield  {author} {\bibinfo {author} {\bibfnamefont {G.~P.}\ \bibnamefont
  {Korchemsky}}, \bibinfo {author} {\bibfnamefont {E.}~\bibnamefont
  {Sokatchev}}, \ and\ \bibinfo {author} {\bibfnamefont {A.}~\bibnamefont
  {Zhiboedov}},\ }\href {\doibase 10.1007/JHEP08(2022)188} {\bibfield
  {journal} {\bibinfo  {journal} {JHEP}\ }\textbf {\bibinfo {volume} {08}},\
  \bibinfo {pages} {188} (\bibinfo {year} {2022})},\ \Eprint
  {http://arxiv.org/abs/2106.14899} {arXiv:2106.14899 [hep-th]} \BibitemShut
  {NoStop}%
\bibitem [{\citenamefont {Korchemsky}\ and\ \citenamefont
  {Zhiboedov}(2022)}]{Korchemsky:2021htm}%
  \BibitemOpen
  \bibfield  {author} {\bibinfo {author} {\bibfnamefont {G.~P.}\ \bibnamefont
  {Korchemsky}}\ and\ \bibinfo {author} {\bibfnamefont {A.}~\bibnamefont
  {Zhiboedov}},\ }\href {\doibase 10.1007/JHEP02(2022)140} {\bibfield
  {journal} {\bibinfo  {journal} {JHEP}\ }\textbf {\bibinfo {volume} {02}},\
  \bibinfo {pages} {140} (\bibinfo {year} {2022})},\ \Eprint
  {http://arxiv.org/abs/2109.13269} {arXiv:2109.13269 [hep-th]} \BibitemShut
  {NoStop}%
\bibitem [{\citenamefont {Chang}\ and\ \citenamefont
  {Simmons-Duffin}(2022)}]{Chang:2022ryc}%
  \BibitemOpen
  \bibfield  {author} {\bibinfo {author} {\bibfnamefont {C.-H.}\ \bibnamefont
  {Chang}}\ and\ \bibinfo {author} {\bibfnamefont {D.}~\bibnamefont
  {Simmons-Duffin}},\ }\href@noop {} {\  (\bibinfo {year} {2022})},\ \Eprint
  {http://arxiv.org/abs/2202.04090} {arXiv:2202.04090 [hep-th]} \BibitemShut
  {NoStop}%
\bibitem [{\citenamefont {Chen}\ \emph
  {et~al.}(2022{\natexlab{b}})\citenamefont {Chen}, \citenamefont {Moult},
  \citenamefont {Sandor},\ and\ \citenamefont {Zhu}}]{Chen:2022jhb}%
  \BibitemOpen
  \bibfield  {author} {\bibinfo {author} {\bibfnamefont {H.}~\bibnamefont
  {Chen}}, \bibinfo {author} {\bibfnamefont {I.}~\bibnamefont {Moult}},
  \bibinfo {author} {\bibfnamefont {J.}~\bibnamefont {Sandor}}, \ and\ \bibinfo
  {author} {\bibfnamefont {H.~X.}\ \bibnamefont {Zhu}},\ }\href@noop {} {\
  (\bibinfo {year} {2022}{\natexlab{b}})},\ \Eprint
  {http://arxiv.org/abs/2202.04085} {arXiv:2202.04085 [hep-ph]} \BibitemShut
  {NoStop}%
\bibitem [{\citenamefont {Chen}\ \emph
  {et~al.}(2022{\natexlab{c}})\citenamefont {Chen}, \citenamefont {Moult},
  \citenamefont {Thaler},\ and\ \citenamefont {Zhu}}]{Chen:2022swd}%
  \BibitemOpen
  \bibfield  {author} {\bibinfo {author} {\bibfnamefont {H.}~\bibnamefont
  {Chen}}, \bibinfo {author} {\bibfnamefont {I.}~\bibnamefont {Moult}},
  \bibinfo {author} {\bibfnamefont {J.}~\bibnamefont {Thaler}}, \ and\ \bibinfo
  {author} {\bibfnamefont {H.~X.}\ \bibnamefont {Zhu}},\ }\href@noop {} {\
  (\bibinfo {year} {2022}{\natexlab{c}})},\ \Eprint
  {http://arxiv.org/abs/2205.02857} {arXiv:2205.02857 [hep-ph]} \BibitemShut
  {NoStop}%
\bibitem [{\citenamefont {Lee}\ \emph {et~al.}(2022)\citenamefont {Lee},
  \citenamefont {Me\c{c}aj},\ and\ \citenamefont {Moult}}]{Lee:2022ige}%
  \BibitemOpen
  \bibfield  {author} {\bibinfo {author} {\bibfnamefont {K.}~\bibnamefont
  {Lee}}, \bibinfo {author} {\bibfnamefont {B.}~\bibnamefont {Me\c{c}aj}}, \
  and\ \bibinfo {author} {\bibfnamefont {I.}~\bibnamefont {Moult}},\
  }\href@noop {} {\  (\bibinfo {year} {2022})},\ \Eprint
  {http://arxiv.org/abs/2205.03414} {arXiv:2205.03414 [hep-ph]} \BibitemShut
  {NoStop}%
\bibitem [{\citenamefont {Yan}\ and\ \citenamefont
  {Zhang}(2022)}]{Yan:2022cye}%
  \BibitemOpen
  \bibfield  {author} {\bibinfo {author} {\bibfnamefont {K.}~\bibnamefont
  {Yan}}\ and\ \bibinfo {author} {\bibfnamefont {X.}~\bibnamefont {Zhang}},\
  }\href@noop {} {\  (\bibinfo {year} {2022})},\ \Eprint
  {http://arxiv.org/abs/2203.04349} {arXiv:2203.04349 [hep-th]} \BibitemShut
  {NoStop}%
\bibitem [{\citenamefont {Yang}\ and\ \citenamefont
  {Zhang}(2022)}]{Yang:2022tgm}%
  \BibitemOpen
  \bibfield  {author} {\bibinfo {author} {\bibfnamefont {T.-Z.}\ \bibnamefont
  {Yang}}\ and\ \bibinfo {author} {\bibfnamefont {X.}~\bibnamefont {Zhang}},\
  }\href {\doibase 10.1007/JHEP09(2022)006} {\bibfield  {journal} {\bibinfo
  {journal} {JHEP}\ }\textbf {\bibinfo {volume} {09}},\ \bibinfo {pages} {006}
  (\bibinfo {year} {2022})},\ \Eprint {http://arxiv.org/abs/2208.01051}
  {arXiv:2208.01051 [hep-ph]} \BibitemShut {NoStop}%
\bibitem [{\citenamefont {Chen}\ \emph {et~al.}(2023)\citenamefont {Chen},
  \citenamefont {Zhou},\ and\ \citenamefont {Zhu}}]{Chen:2023wah}%
  \BibitemOpen
  \bibfield  {author} {\bibinfo {author} {\bibfnamefont {H.}~\bibnamefont
  {Chen}}, \bibinfo {author} {\bibfnamefont {X.}~\bibnamefont {Zhou}}, \ and\
  \bibinfo {author} {\bibfnamefont {H.~X.}\ \bibnamefont {Zhu}},\ }\href@noop
  {} {\  (\bibinfo {year} {2023})},\ \Eprint {http://arxiv.org/abs/2301.03616}
  {arXiv:2301.03616 [hep-ph]} \BibitemShut {NoStop}%
\bibitem [{\citenamefont {Komiske}\ \emph {et~al.}(2023)\citenamefont
  {Komiske}, \citenamefont {Moult}, \citenamefont {Thaler},\ and\ \citenamefont
  {Zhu}}]{Komiske:2022enw}%
  \BibitemOpen
  \bibfield  {author} {\bibinfo {author} {\bibfnamefont {P.~T.}\ \bibnamefont
  {Komiske}}, \bibinfo {author} {\bibfnamefont {I.}~\bibnamefont {Moult}},
  \bibinfo {author} {\bibfnamefont {J.}~\bibnamefont {Thaler}}, \ and\ \bibinfo
  {author} {\bibfnamefont {H.~X.}\ \bibnamefont {Zhu}},\ }\href {\doibase
  10.1103/PhysRevLett.130.051901} {\bibfield  {journal} {\bibinfo  {journal}
  {Phys. Rev. Lett.}\ }\textbf {\bibinfo {volume} {130}},\ \bibinfo {pages}
  {051901} (\bibinfo {year} {2023})},\ \Eprint
  {http://arxiv.org/abs/2201.07800} {arXiv:2201.07800 [hep-ph]} \BibitemShut
  {NoStop}%
\bibitem [{\citenamefont {Holguin}\ \emph {et~al.}(2022)\citenamefont
  {Holguin}, \citenamefont {Moult}, \citenamefont {Pathak},\ and\ \citenamefont
  {Procura}}]{Holguin:2022epo}%
  \BibitemOpen
  \bibfield  {author} {\bibinfo {author} {\bibfnamefont {J.}~\bibnamefont
  {Holguin}}, \bibinfo {author} {\bibfnamefont {I.}~\bibnamefont {Moult}},
  \bibinfo {author} {\bibfnamefont {A.}~\bibnamefont {Pathak}}, \ and\ \bibinfo
  {author} {\bibfnamefont {M.}~\bibnamefont {Procura}},\ }\href@noop {} {\
  (\bibinfo {year} {2022})},\ \Eprint {http://arxiv.org/abs/2201.08393}
  {arXiv:2201.08393 [hep-ph]} \BibitemShut {NoStop}%
\bibitem [{\citenamefont {Liu}\ and\ \citenamefont {Zhu}(2023)}]{Liu:2022wop}%
  \BibitemOpen
  \bibfield  {author} {\bibinfo {author} {\bibfnamefont {X.}~\bibnamefont
  {Liu}}\ and\ \bibinfo {author} {\bibfnamefont {H.~X.}\ \bibnamefont {Zhu}},\
  }\href {\doibase 10.1103/PhysRevLett.130.091901} {\bibfield  {journal}
  {\bibinfo  {journal} {Phys. Rev. Lett.}\ }\textbf {\bibinfo {volume} {130}},\
  \bibinfo {pages} {091901} (\bibinfo {year} {2023})},\ \Eprint
  {http://arxiv.org/abs/2209.02080} {arXiv:2209.02080 [hep-ph]} \BibitemShut
  {NoStop}%
\bibitem [{\citenamefont {Liu}\ \emph {et~al.}(2023)\citenamefont {Liu},
  \citenamefont {Liu}, \citenamefont {Pan}, \citenamefont {Yuan},\ and\
  \citenamefont {Zhu}}]{Liu:2023aqb}%
  \BibitemOpen
  \bibfield  {author} {\bibinfo {author} {\bibfnamefont {H.-Y.}\ \bibnamefont
  {Liu}}, \bibinfo {author} {\bibfnamefont {X.}~\bibnamefont {Liu}}, \bibinfo
  {author} {\bibfnamefont {J.-C.}\ \bibnamefont {Pan}}, \bibinfo {author}
  {\bibfnamefont {F.}~\bibnamefont {Yuan}}, \ and\ \bibinfo {author}
  {\bibfnamefont {H.~X.}\ \bibnamefont {Zhu}},\ }\href@noop {} {\  (\bibinfo
  {year} {2023})},\ \Eprint {http://arxiv.org/abs/2301.01788} {arXiv:2301.01788
  [hep-ph]} \BibitemShut {NoStop}%
\bibitem [{\citenamefont {Cao}\ \emph {et~al.}(2023)\citenamefont {Cao},
  \citenamefont {Liu},\ and\ \citenamefont {Zhu}}]{Cao:2023rga}%
  \BibitemOpen
  \bibfield  {author} {\bibinfo {author} {\bibfnamefont {H.}~\bibnamefont
  {Cao}}, \bibinfo {author} {\bibfnamefont {X.}~\bibnamefont {Liu}}, \ and\
  \bibinfo {author} {\bibfnamefont {H.~X.}\ \bibnamefont {Zhu}},\ }\href@noop
  {} {\  (\bibinfo {year} {2023})},\ \Eprint {http://arxiv.org/abs/2303.01530}
  {arXiv:2303.01530 [hep-ph]} \BibitemShut {NoStop}%
\bibitem [{\citenamefont {Devereaux}\ \emph {et~al.}(2023)\citenamefont
  {Devereaux}, \citenamefont {Fan}, \citenamefont {Ke}, \citenamefont {Lee},\
  and\ \citenamefont {Moult}}]{Devereaux:2023vjz}%
  \BibitemOpen
  \bibfield  {author} {\bibinfo {author} {\bibfnamefont {K.}~\bibnamefont
  {Devereaux}}, \bibinfo {author} {\bibfnamefont {W.}~\bibnamefont {Fan}},
  \bibinfo {author} {\bibfnamefont {W.}~\bibnamefont {Ke}}, \bibinfo {author}
  {\bibfnamefont {K.}~\bibnamefont {Lee}}, \ and\ \bibinfo {author}
  {\bibfnamefont {I.}~\bibnamefont {Moult}},\ }\href@noop {} {\  (\bibinfo
  {year} {2023})},\ \Eprint {http://arxiv.org/abs/2303.08143} {arXiv:2303.08143
  [hep-ph]} \BibitemShut {NoStop}%
\bibitem [{\citenamefont {Andres}\ \emph
  {et~al.}(2023{\natexlab{a}})\citenamefont {Andres}, \citenamefont
  {Dominguez}, \citenamefont {Kunnawalkam~Elayavalli}, \citenamefont {Holguin},
  \citenamefont {Marquet},\ and\ \citenamefont {Moult}}]{Andres:2022ovj}%
  \BibitemOpen
  \bibfield  {author} {\bibinfo {author} {\bibfnamefont {C.}~\bibnamefont
  {Andres}}, \bibinfo {author} {\bibfnamefont {F.}~\bibnamefont {Dominguez}},
  \bibinfo {author} {\bibfnamefont {R.}~\bibnamefont {Kunnawalkam~Elayavalli}},
  \bibinfo {author} {\bibfnamefont {J.}~\bibnamefont {Holguin}}, \bibinfo
  {author} {\bibfnamefont {C.}~\bibnamefont {Marquet}}, \ and\ \bibinfo
  {author} {\bibfnamefont {I.}~\bibnamefont {Moult}},\ }\href {\doibase
  10.1103/PhysRevLett.130.262301} {\bibfield  {journal} {\bibinfo  {journal}
  {Phys. Rev. Lett.}\ }\textbf {\bibinfo {volume} {130}},\ \bibinfo {pages}
  {262301} (\bibinfo {year} {2023}{\natexlab{a}})},\ \Eprint
  {http://arxiv.org/abs/2209.11236} {arXiv:2209.11236 [hep-ph]} \BibitemShut
  {NoStop}%
\bibitem [{\citenamefont {Andres}\ \emph
  {et~al.}(2023{\natexlab{b}})\citenamefont {Andres}, \citenamefont
  {Dominguez}, \citenamefont {Holguin}, \citenamefont {Marquet},\ and\
  \citenamefont {Moult}}]{Andres:2023xwr}%
  \BibitemOpen
  \bibfield  {author} {\bibinfo {author} {\bibfnamefont {C.}~\bibnamefont
  {Andres}}, \bibinfo {author} {\bibfnamefont {F.}~\bibnamefont {Dominguez}},
  \bibinfo {author} {\bibfnamefont {J.}~\bibnamefont {Holguin}}, \bibinfo
  {author} {\bibfnamefont {C.}~\bibnamefont {Marquet}}, \ and\ \bibinfo
  {author} {\bibfnamefont {I.}~\bibnamefont {Moult}},\ }\href@noop {} {\
  (\bibinfo {year} {2023}{\natexlab{b}})},\ \Eprint
  {http://arxiv.org/abs/2303.03413} {arXiv:2303.03413 [hep-ph]} \BibitemShut
  {NoStop}%
\bibitem [{\citenamefont {Craft}\ \emph {et~al.}(2022)\citenamefont {Craft},
  \citenamefont {Lee}, \citenamefont {Me\c{c}aj},\ and\ \citenamefont
  {Moult}}]{Craft:2022kdo}%
  \BibitemOpen
  \bibfield  {author} {\bibinfo {author} {\bibfnamefont {E.}~\bibnamefont
  {Craft}}, \bibinfo {author} {\bibfnamefont {K.}~\bibnamefont {Lee}}, \bibinfo
  {author} {\bibfnamefont {B.}~\bibnamefont {Me\c{c}aj}}, \ and\ \bibinfo
  {author} {\bibfnamefont {I.}~\bibnamefont {Moult}},\ }\href@noop {} {\
  (\bibinfo {year} {2022})},\ \Eprint {http://arxiv.org/abs/2210.09311}
  {arXiv:2210.09311 [hep-ph]} \BibitemShut {NoStop}%
\bibitem [{\citenamefont {Dom\'\i{}nguez}\ \emph {et~al.}(2020)\citenamefont
  {Dom\'\i{}nguez}, \citenamefont {Milhano}, \citenamefont {Salgado},
  \citenamefont {Tywoniuk},\ and\ \citenamefont {Vila}}]{Dominguez:2019ges}%
  \BibitemOpen
  \bibfield  {author} {\bibinfo {author} {\bibfnamefont {F.}~\bibnamefont
  {Dom\'\i{}nguez}}, \bibinfo {author} {\bibfnamefont {J.~G.}\ \bibnamefont
  {Milhano}}, \bibinfo {author} {\bibfnamefont {C.~A.}\ \bibnamefont
  {Salgado}}, \bibinfo {author} {\bibfnamefont {K.}~\bibnamefont {Tywoniuk}}, \
  and\ \bibinfo {author} {\bibfnamefont {V.}~\bibnamefont {Vila}},\ }\href
  {\doibase 10.1140/epjc/s10052-019-7563-0} {\bibfield  {journal} {\bibinfo
  {journal} {Eur. Phys. J. C}\ }\textbf {\bibinfo {volume} {80}},\ \bibinfo
  {pages} {11} (\bibinfo {year} {2020})},\ \Eprint
  {http://arxiv.org/abs/1907.03653} {arXiv:1907.03653 [hep-ph]} \BibitemShut
  {NoStop}%
\bibitem [{\citenamefont {Isaksen}\ and\ \citenamefont
  {Tywoniuk}(2020)}]{Isaksen:2020npj}%
  \BibitemOpen
  \bibfield  {author} {\bibinfo {author} {\bibfnamefont {J.~H.}\ \bibnamefont
  {Isaksen}}\ and\ \bibinfo {author} {\bibfnamefont {K.}~\bibnamefont
  {Tywoniuk}},\ }\href {\doibase 10.1007/JHEP11(2021)125} {\bibfield  {journal}
  {\bibinfo  {journal} {JHEP}\ }\textbf {\bibinfo {volume} {21}},\ \bibinfo
  {pages} {125} (\bibinfo {year} {2020})},\ \Eprint
  {http://arxiv.org/abs/2107.02542} {arXiv:2107.02542 [hep-ph]} \BibitemShut
  {NoStop}%
\bibitem [{\citenamefont {Baier}\ \emph
  {et~al.}(1997{\natexlab{a}})\citenamefont {Baier}, \citenamefont
  {Dokshitzer}, \citenamefont {Mueller}, \citenamefont {Peigne},\ and\
  \citenamefont {Schiff}}]{Baier:1996kr}%
  \BibitemOpen
  \bibfield  {author} {\bibinfo {author} {\bibfnamefont {R.}~\bibnamefont
  {Baier}}, \bibinfo {author} {\bibfnamefont {Y.~L.}\ \bibnamefont
  {Dokshitzer}}, \bibinfo {author} {\bibfnamefont {A.~H.}\ \bibnamefont
  {Mueller}}, \bibinfo {author} {\bibfnamefont {S.}~\bibnamefont {Peigne}}, \
  and\ \bibinfo {author} {\bibfnamefont {D.}~\bibnamefont {Schiff}},\ }\href
  {\doibase 10.1016/S0550-3213(96)00553-6} {\bibfield  {journal} {\bibinfo
  {journal} {Nucl. Phys. B}\ }\textbf {\bibinfo {volume} {483}},\ \bibinfo
  {pages} {291} (\bibinfo {year} {1997}{\natexlab{a}})},\ \Eprint
  {http://arxiv.org/abs/hep-ph/9607355} {arXiv:hep-ph/9607355} \BibitemShut
  {NoStop}%
\bibitem [{\citenamefont {Baier}\ \emph
  {et~al.}(1997{\natexlab{b}})\citenamefont {Baier}, \citenamefont
  {Dokshitzer}, \citenamefont {Mueller}, \citenamefont {Peigne},\ and\
  \citenamefont {Schiff}}]{Baier:1996sk}%
  \BibitemOpen
  \bibfield  {author} {\bibinfo {author} {\bibfnamefont {R.}~\bibnamefont
  {Baier}}, \bibinfo {author} {\bibfnamefont {Y.~L.}\ \bibnamefont
  {Dokshitzer}}, \bibinfo {author} {\bibfnamefont {A.~H.}\ \bibnamefont
  {Mueller}}, \bibinfo {author} {\bibfnamefont {S.}~\bibnamefont {Peigne}}, \
  and\ \bibinfo {author} {\bibfnamefont {D.}~\bibnamefont {Schiff}},\ }\href
  {\doibase 10.1016/S0550-3213(96)00581-0} {\bibfield  {journal} {\bibinfo
  {journal} {Nucl. Phys. B}\ }\textbf {\bibinfo {volume} {484}},\ \bibinfo
  {pages} {265} (\bibinfo {year} {1997}{\natexlab{b}})},\ \Eprint
  {http://arxiv.org/abs/hep-ph/9608322} {arXiv:hep-ph/9608322} \BibitemShut
  {NoStop}%
\bibitem [{\citenamefont {Zakharov}(1996)}]{Zakharov:1996fv}%
  \BibitemOpen
  \bibfield  {author} {\bibinfo {author} {\bibfnamefont {B.~G.}\ \bibnamefont
  {Zakharov}},\ }\href {\doibase 10.1134/1.567126} {\bibfield  {journal}
  {\bibinfo  {journal} {JETP Lett.}\ }\textbf {\bibinfo {volume} {63}},\
  \bibinfo {pages} {952} (\bibinfo {year} {1996})},\ \Eprint
  {http://arxiv.org/abs/hep-ph/9607440} {arXiv:hep-ph/9607440} \BibitemShut
  {NoStop}%
\bibitem [{\citenamefont {Zakharov}(1997)}]{Zakharov:1997uu}%
  \BibitemOpen
  \bibfield  {author} {\bibinfo {author} {\bibfnamefont {B.~G.}\ \bibnamefont
  {Zakharov}},\ }\href {\doibase 10.1134/1.567389} {\bibfield  {journal}
  {\bibinfo  {journal} {JETP Lett.}\ }\textbf {\bibinfo {volume} {65}},\
  \bibinfo {pages} {615} (\bibinfo {year} {1997})},\ \Eprint
  {http://arxiv.org/abs/hep-ph/9704255} {arXiv:hep-ph/9704255} \BibitemShut
  {NoStop}%
\bibitem [{\citenamefont {Gyulassy}\ and\ \citenamefont
  {Wang}(1994)}]{Gyulassy:1993hr}%
  \BibitemOpen
  \bibfield  {author} {\bibinfo {author} {\bibfnamefont {M.}~\bibnamefont
  {Gyulassy}}\ and\ \bibinfo {author} {\bibfnamefont {X.-n.}\ \bibnamefont
  {Wang}},\ }\href {\doibase 10.1016/0550-3213(94)90079-5} {\bibfield
  {journal} {\bibinfo  {journal} {Nucl. Phys. B}\ }\textbf {\bibinfo {volume}
  {420}},\ \bibinfo {pages} {583} (\bibinfo {year} {1994})},\ \Eprint
  {http://arxiv.org/abs/nucl-th/9306003} {arXiv:nucl-th/9306003} \BibitemShut
  {NoStop}%
\bibitem [{\citenamefont {Catani}\ \emph {et~al.}(2001)\citenamefont {Catani},
  \citenamefont {Dittmaier},\ and\ \citenamefont {Trocsanyi}}]{Catani:2000ef}%
  \BibitemOpen
  \bibfield  {author} {\bibinfo {author} {\bibfnamefont {S.}~\bibnamefont
  {Catani}}, \bibinfo {author} {\bibfnamefont {S.}~\bibnamefont {Dittmaier}}, \
  and\ \bibinfo {author} {\bibfnamefont {Z.}~\bibnamefont {Trocsanyi}},\ }\href
  {\doibase 10.1016/S0370-2693(01)00065-X} {\bibfield  {journal} {\bibinfo
  {journal} {Phys. Lett. B}\ }\textbf {\bibinfo {volume} {500}},\ \bibinfo
  {pages} {149} (\bibinfo {year} {2001})},\ \Eprint
  {http://arxiv.org/abs/hep-ph/0011222} {arXiv:hep-ph/0011222} \BibitemShut
  {NoStop}%
\bibitem [{\citenamefont {Barata}\ and\ \citenamefont
  {Mehtar-Tani}(2023)}]{Barata:2023vnl}%
  \BibitemOpen
  \bibfield  {author} {\bibinfo {author} {\bibfnamefont {J.~a.}\ \bibnamefont
  {Barata}}\ and\ \bibinfo {author} {\bibfnamefont {Y.}~\bibnamefont
  {Mehtar-Tani}}\ }(\bibinfo {year} {2023})\ \Eprint
  {http://arxiv.org/abs/2307.08943} {arXiv:2307.08943 [hep-ph]} \BibitemShut
  {NoStop}%
\bibitem [{\citenamefont {Fan}()}]{talk1}%
  \BibitemOpen
  \bibfield  {author} {\bibinfo {author} {\bibfnamefont {W.}~\bibnamefont
  {Fan}},\ }\href@noop {} {\bibinfo  {journal} {Imaging Cold Nuclear Matter
  with Energy Correlators at the future EIC, DIS 2023}\ }\BibitemShut {NoStop}%
\bibitem [{\citenamefont {Cruz-Torres}()}]{talk2}%
  \BibitemOpen
\bibfield  {journal} {  }\bibfield  {author} {\bibinfo {author} {\bibfnamefont
  {R.}~\bibnamefont {Cruz-Torres}},\ }\href@noop {} {\bibinfo  {journal}
  {Measurement of the angle between jet axes and energy-energy correlators with
  ALICE, Hard Probes 2023}\ }\BibitemShut {NoStop}%
\bibitem [{\citenamefont {Tamis}()}]{talk3}%
  \BibitemOpen
\bibfield  {journal} {  }\bibfield  {author} {\bibinfo {author} {\bibfnamefont
  {A.}~\bibnamefont {Tamis}},\ }\href@noop {} {\bibinfo  {journal} {Measurement
  of Two-Point Energy Correlators Within Jets in p+p Collisions at $\sqrt{s}$ =
  200 GeV, Hard Probes 2023}\ }\BibitemShut {NoStop}%
\bibitem [{\citenamefont {Sievert}\ and\ \citenamefont
  {Vitev}(2018)}]{Sievert:2018imd}%
  \BibitemOpen
\bibfield  {journal} {  }\bibfield  {author} {\bibinfo {author} {\bibfnamefont
  {M.~D.}\ \bibnamefont {Sievert}}\ and\ \bibinfo {author} {\bibfnamefont
  {I.}~\bibnamefont {Vitev}},\ }\href {\doibase 10.1103/PhysRevD.98.094010}
  {\bibfield  {journal} {\bibinfo  {journal} {Phys. Rev. D}\ }\textbf {\bibinfo
  {volume} {98}},\ \bibinfo {pages} {094010} (\bibinfo {year} {2018})},\
  \Eprint {http://arxiv.org/abs/1807.03799} {arXiv:1807.03799 [hep-ph]}
  \BibitemShut {NoStop}%
\end{thebibliography}%
\bibliographystyle{apsrev4-1}
\newpage
\onecolumngrid
\newpage
%%%%%%%%%%%%%%%%%%%%%%%%%%%%%%%%%%%%%%%%%%%%%%%%%%%%%%%%%%%%%%%%%%%%%%%%%%%%

\end{document}